# Disordered Optical Metasurfaces: Basics, Properties, and Applications

Philippe Lalanne[1,*], Miao Chen[1], Carsten Rockstuhl,[2,3] Alexander Sprafke,[4] Alexandre Dmitriev,[5] and Kevin Vynck[6]

[1]*Laboratoire Photonique Numérique et Nanosciences (LP2N), Université de Bordeaux, Institut d'Optique Graduate School, CNRS, Talence, France*
[2]*Institute of Theoretical Solid State Physics, Karlsruhe Institute of Technology, 76131 Karlsruhe, Germany*
[3]*Institute of Nanotechnology, Karlsruhe Institute of Technology, 76131 Karlsruhe, Germany*
[4]*Institute of Physics, Martin Luther University Halle-Wittenberg, 06120 Halle, Germany*
[5]*Department of Physics, University of Gothenburg, 41296 Gothenburg, Sweden*
[6]*Institut Lumière Matière (iLM), Université Claude Bernard Lyon 1, CNRS, Villeurbanne, France*
*philippe.lalanne@institutoptique.fr*

**Abstract:** Optical metasurfaces are conventionally viewed as organized flat arrays of photonic or plasmonic nanoresonators, also called metaatoms. These metasurfaces are typically highly ordered and fabricated with precision using expensive tools. However, the inherent imperfections in large-scale nanophotonic devices, along with recent advances in bottom-up nanofabrication techniques and design strategies, have highlighted the potential benefits of incorporating disorder to achieve specific optical functionalities. This review offers an overview of the key theoretical, numerical, and experimental aspects related to the exploration of disordered optical metasurfaces. It introduces fundamental concepts of light scattering by disordered metasurfaces and outlines theoretical and numerical methodologies for analyzing their optical behavior. Various fabrication techniques are discussed, highlighting the types of disorder they deliver and their achievable precision level. The review also explores critical applications of disordered optical metasurfaces, such as light manipulation in thin film materials and the design of structural colors and visual appearances. Finally, the article offers perspectives on the burgeoning future research in this field. Disordered optical metasurfaces offer a promising alternative to their ordered counterparts, often delivering unique functionalities or enhanced performance. They present a particularly exciting opportunity in applications demanding large-scale implementation, such as sustainable renewable energy systems, as well as aesthetically vibrant coatings for luxury goods and architectural designs.

## 1. Introduction

The term "optical metasurfaces" encompasses a wide range of nanostructured surfaces, where the constituent scatterers (referred to as metaatoms) are usually crafted through precise engineering of high-index dielectric or metallic layers. Disordered optical metasurfaces represent a specific subset within this category, characterized by randomly positioned metaatoms. The term 'metaatom' is derived from the idea that, just as atoms are the building blocks of conventional materials, metaatoms serve as the foundational units of metamaterials or metasurfaces and are designed to interact with electromagnetic waves in ways that bulk materials cannot. The primary objective of this review is to comprehend the optical properties of statistically homogeneous collections of metaatoms, achieved through averaging the optical response over multiple independent instances of disorder [1-3].

There are several compelling reasons to study disordered optical metasurfaces, even for researchers primarily focused on ordered ones. First, irregularities or imperfections, although seemingly absent in ordered metasurfaces, can still exert significant influence on the optical properties [4]. Moreover, disordered metasurfaces can be manufactured over large areas at a low cost, facilitating scalable production. Nature presents numerous intriguing examples of disordered metasurfaces, showcasing a wide range of shapes and arrangements that captivate our interest [5,6]. The study of disordered metasurfaces also unveils a rich spectrum of physics, ranging from nanoscale resonances to wavelength-scale mode hybridization and collective interference phenomena at the mesoscale. Lastly, akin to ordered metasurfaces, disordered metasurfaces have a plethora of identified applications, some predating the term "metasurface," such as anti-reflection coatings [7], thin-film photovoltaic cells [8], low-emissivity coatings [9], surface-enhanced Raman spectroscopy (SERS) sensors [10], among others discussed later.

The study of disordered optical surfaces amalgamates knowledge of wave transport in complex media with insights from collective Mie and plasmonic resonances typically observed in ordered metasurfaces. While possessing a solid understanding in both domains before exploring disordered metasurfaces is beneficial, in-depth expertise is not a prerequisite. This is because numerous advanced and specific concepts pertinent to disordered media or ordered metasurfaces may not directly apply in the present context. For instance, the concept of a mean free path, crucial for distinguishing between the ballistic and diffusive regimes in bulk turbid media, loses its relevance for metasurfaces. Additionally, periodicity and associated theoretical constructs like Bloch modes and bandgaps, commonly encountered in ordered metasurfaces, are absent in disordered counterparts.

Building on this observation, the review is organized to ensure accessibility with minimal background knowledge on disordered media or ordered metasurfaces. We aim to achieve this ambitious goal by systematically correlating basic concepts with their practical applications. By navigating between the fundamental content in Sections 2,3 and 4 and the applied content in Sections 5 and 6, readers can attain a high level of comprehension with minimal prior knowledge.

The current review mostly focuses on metasurfaces featuring clearly defined scatterers, such as inclusions or metaatoms. Despite the multitude of applications [11,12], the richness of optical responses, and the vast array of fabrication options [13], the review weakly cover metasurfaces consisting in random nanoparticle clusters or two-phase composites with arbitrary microstructures [13,14]. The theoretical analysis of such dense configurations must carefully consider capacitive coupling and spatial dispersion [15] and attaining a thorough comprehension of these systems remains challenging, given the current state of knowledge.

Previous literature on this topic is limited, with only two known reviews [16,17]. Our current review distinguishes itself from these predecessors. In [16], a thorough examination of the applications of disordered metasurfaces is provided. In contrast, our focus leans more heavily towards elucidating fundamental concepts, with a less exhaustive treatment of applications. We aim to illustrate how these fundamental principles can be practically implemented, showcasing the utilization of degrees of freedom to design specific functions. The review presented in [17], authored by almost the same individuals as the current one, shares a similar approach. However, the present work addresses the subject in considerably more details, although we intentionally refrain from exhaustive coverage to highlight the foundational aspects.

The Review is divided into seven additional sections.

Section 2 presents an attempt to establish a metric for classifying metasurfaces based on their levels of order or disorder. Such a classification is crucial for designing metasurfaces with specific scattering properties.

Section 3 presents key concepts regarding disordered media. We first introduce the concepts of specular (a.k.a. coherent) and diffuse (a.k.a. incoherent) components of the scattered light by employing statistical ensemble averages to discriminate the two quantities. Subsequently, we introduce the Bidirectional Scattering Distribution Function (BSDF), a pivotal radiometric measure quantifying how random surfaces, statistically invariant by translation, scatter light into these specular and diffuse components.

Section 4 covers state-of-the-art numerical techniques that address the outstanding challenge of predicting the optical properties of finite disordered metasurfaces. Emphasis is placed on the techniques that facilitate the simulation of large collections of metaatoms, highlighting the challenge of extrapolating the optical properties of metasurfaces with infinite lateral dimensions from finite-size numerical simulations.

Section 5 reviews several semi-analytical models that can predict the diffuse and specular contributions of the BSDF for infinite metasurfaces. These models are powerful as they reveal the control parameters for manipulating the scattering properties of metasurfaces angularly and spectrally, aiding in design studies.

Section 6 focuses on the rapidly advancing capabilities in nanofabrication, highlighting approaches adapted for manufacturing disordered optical metasurfaces. Unlike conventional tools used for fabricating ordered metasurfaces with subwavelength-scale patterns, such as phase-shift or deep UV lithography, we focus on bottom-up approaches that hold promise for achieving cost-effective industrial-scale production over extensive areas.

Section 7 thoroughly examines various important applications of disordered optical metasurfaces, emphasizing current cutting-edge research and future prospect. Key functionalities and devices are discussed, including diffusers, absorbers, extractors, and transparent displays. We methodically demonstrate how insights gained from the models discussed in Section 5 can help designers to achieve devices with significantly improved performance.

Section 8 summarizes the Review and provides some perspectives for future work.

## 2. Classification of disordered metasurfaces

Many degrees of freedom may be exploited to design disordered metasurfaces, such as metaatom positions, shapes, sizes, and materials. One of the main goals of designers is to classify disordered metasurfaces using metrics that quantify their degrees of order or disorder. Doing so, they can devise experimental or numerical protocols that generate metasurfaces with prescribed disorder levels and link these levels to desirable physical properties [18,19]. Additionally, such a classification could facilitate a deeper understanding of the relative advantages and disadvantages of disordered versus periodic metasurfaces. However, devising comprehensive classifications and metrics is challenging due to the immense variability in metaatom arrangements and morphologies (see Fig. 1). The challenge lies in formulating metrics that capture order across different spatial scales: short-, intermediate-, and long-range [18-19,26].

Current research has primarily focused on particulate metasurfaces, which are composed of distinct, individual metaatoms. By conceptualizing these metaatoms as points, n-point correlation functions emerge as the best-suited metric for classifying morphologies, especially when lying between perfect periodicity and complete randomness. The 2-point correlation function, formally related to the structure factor discussed in Section 5.2, is the most widely used among these in the literature, as it gives the leading contribution of structural correlations on the scattered intensity [19]. The impact of higher-order correlations is more difficult to apprehend. Interested readers are referred to advanced studies (e.g., Ref. [18]) for a deeper exploration of these metrics.

Figure 1a offers a reduced classification, fully sufficient for our purpose, of particulate morphologies common in photonics, arranged along a spectrum from perfect order, i.e. crystals (Fig. 1a1), to uncorrelated disorder, i.e., random arrangements (Fig. 1a5). Crystals exhibit both long-range translational and rotational order around specific points in space, producing well-defined diffraction patterns in momentum space. Quasicrystals (not shown) lack translational but retain rotational order around specific points in space [27]. A quasicrystal is a structure that achieves complete surface filling through an aperiodic arrangement of multiple distinct basic tiling, unlike periodic crystals that use a single repeating unit. At the other end, uncorrelated disorder (Fig. 1a5), typically generated by a Poisson point process (i.e., placing points at fully random positions), exhibit statistical translational and rotational invariances and lead to a white noise spectrum in momentum space. Albeit convenient for theoretical models, real nanostructures always exhibit a certain degree of spatial correlation.

Intermediate states on this 2D order-disorder spectrum include imperfect crystals (Fig. 1a2), stealthy hyperuniform disordered structures (Fig. 1a3), and amorphous (i.e. short-range correlated disordered) structures (Fig. 1a4). Each disorder state generates distinctive scattering profiles, which are discussed in Appendix D.

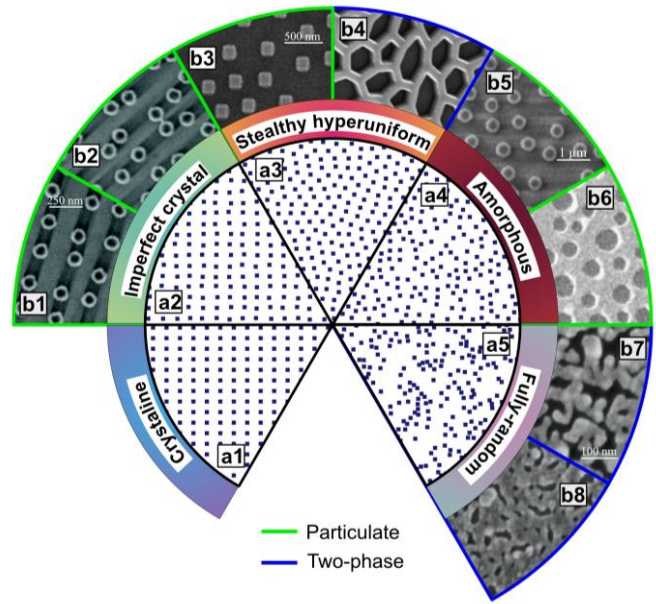

**Figure 1.** Classification of disordered metasurfaces. **(a1)-(a5)** Selected typical states of 2D point-like statistically homogeneous morphologies organized according to their "level" of order, from to highest (crystal) to lowest (fully-random): **(a1)** crystal state; **(a2)** imperfect crystal ; **(a3)** stealthy hyperuniform state ; **(a4)** amorphous state ; **(a5)** fully-random state. **(b1)-(b8)** Real metasurfaces can be divided into particulate and two-phase states. **(b1)-(b2)** Imperfect crystals obtained with small and large random displacements of the metaatoms, respectively. **(b3)** Stealthy hyperuniform arrangement of Si nanoboxes on silica fabricated with e-beam lithography [20]. **(b4)** Disordered hyperuniform metasurfaces fabricated with e-beam lithography for thin-film photovoltaic applications. **(b5)** Amorphous arrangement with $TiO_2$ nanodisks fabricated by colloidal chemistry. **(b6)** Air holes in Ag films fabricated by island lithography. **(b7)-(b8)** Interrupted Ag film deposition at percolation threshold **(b5)** and above **(b6)**. Panel **(b1)-(b2)** used with permission of Royal Society of Chemistry, from "Influence of order-to-disorder transitions on the optical properties of the aluminum plasmonic metasurface." Zhang *et al*., Nanoscale, **12**, 23173-23182 [21], Copyright 2020; permission conveyed through Copyright Clearance Center, Inc. Panel **(b4)** reprinted from [22], CC BY 4.0. Panel **(b5)** reprinted from Piechulla *et al*., Adv. Opt. Mater. **9**, 2100186 (2021) [23]. Copyright Wiley-VCH Verlag GmbH & Co. KGaA. Reproduced with permission. Panel **(b6)** reprinted from Thin Solid Films **467**, Green and Yi, "Light transmission through perforated metal thin films made by island lithography," 308–312 [24], Copyright 2004, with permission from Elsevier. Panel **(b7)-(b8)** used with permission of IOP Publishing, Ltd, from "Optimum plasmon hybridization at percolation threshold of silver films near metallic surfaces." Maaroof *et al.*, J. Phys. D: Appl. Phys, **43**, 405301 [25], Copyright 2010; permission conveyed through Copyright Clearance Center, Inc.

Imperfect crystals are periodic structures disrupted by local random variations, such as fluctuations in the sizes or positions of meta-atoms. Famous examples in crystallography are vacancy or substitutional defects. The imperfect crystal in Fig. 1a2 is obtained by displacing the position of each metaatom, initially located on a square lattice, within their unit cell. Thus, only short-range disorder is introduced in this simple and common method [21], while long-range order is preserved as the structure remains periodic on average. The long-range order makes such imperfect structures preferentially fabricable with electron beam lithography rather than self-assembly approaches.

Hyperuniform structures (Fig. 1a3) achieve an extraordinary degree of spatial uniformity over large scales, despite being disordered on smaller scales. In 2D, the fluctuations in the point density for hyperuniform systems grows as the system size, whereas it grows as square of the size for conventional disordered systems. We will see that hyperuniform metasurfaces do not diffuse light in the specular direction. In even more advanced forms, stealthy hyperuniform structures achieve an even stronger level of uniformity, characterized not only by the absence of diffuse light in the specular direction but also within a finite angular range around it. Achieving this level of control, however, requires precise positioning of the constituent particles, making stealthy hyperuniform structures, like imperfect crystals (Fig. 1a2), challenging to fabricate at large scales using bottom-up methods [23] (Section 6).

Amorphous disorder (Fig. 1a4) exhibits short-range disorder correlations typical of liquids of nonoverlapping spheres or disks. As particle density increases, particles organize to optimize space-filling, resulting in short-range correlations without long-range order. This type of disorder is prevalent in natural micro- and nanostructures, like bird feathers or butterflies [5,6], and is also common in colloidal suspensions, where particle impenetrability generates short-range correlations. Various deposition techniques, such as drop casting, can produce short-range correlated disordered metasurfaces with a broad variety of metaatom by using charged nanoparticles to control interparticle electrostatic repulsion and avoid clustering [11,23,28].

The classification shown in Fig. 1a focuses exclusively on metaatom positions and does not account for variations in metaatom shapes, sizes, or densities. Within the independent scattering approximation, the concept of a structure factor is central (see Section 5), enabling predictions of the optical properties of particulate morphologies composed of identical metaatoms. However, this position-based classification captures only a limited subset of a more extensive design landscape, which is illustrated in Fig. 1b with examples of realistic metasurfaces.

In practical settings, even when starting with a monolayer of identical, well-identified metaatoms that fit into our order classification, multiple factors can influence the metasurface optical properties. For instance, significant variations in shape (Fig. 1b4) or size (Fig. 1b6) may reduce the predictive accuracy of approximate models based on structure-factor approaches.

Additionally, as metaatom size or density increases, overlap between metaatoms can occur, causing a transition from particulate morphologies to two-phase composites. This shift is evident in cases like interrupted metal film deposition, widely studied for SERS applications (Section 6). Initially, low quantities of deposited metal form isolated nano-islands. With additional deposition, these islands grow and merge into larger islands, eventually reaching the percolation threshold [29], where continuous paths appear throughout the sample (Fig. 1b7). With further deposition, metal becomes the dominant phase, and air gaps become rare, resulting in an almost continuous film (Fig. 1b8). At this stage, the initial particulate morphology is replaced by a two-phase composite.

This transition from particulate to two-phase morphologies is associated with significant changes in optical properties. For example, studies on SERS have shown that the near-field characteristics of semicontinuous metal films differ substantially above and below the percolation threshold [30,31]. However, the extent to which particulate-state properties are preserved during this transition remains unclear. This uncertainty partly stems from the scarcity of experimental studies on two-phase metasurfaces and the fact that modeling two-phase and particulate metasurfaces requires fully different numerical approaches.

Two-phase metasurfaces are best studied with general-purpose Maxwell solvers, such as the finite-difference time-domain method, while particulate morphologies are typically analyzed with specialized tools based on electromagnetic Green-tensor approaches (Section 4). This division has led to independent lines of interpretation and conceptualization for each class of metasurfaces. Consequently, this review reflects a predominant focus on theoretical and experimental research related to particulate morphologies, which has historically received more attention.

## 3. Key principles of light scattering by disordered metasurfaces

The problem of scattering by disordered metasurfaces has many similarities with the physics of wave scattering by random rough surfaces. The latter is a long-standing topic covered by several excellent textbooks [1-3] and finds applications in various fields, including radar remote sensing, LiDAR, underwater acoustics, wireless communications, seismology, computer graphics for instance.

When illuminated by a monochromatic planewave, the far-field scattering diagram (or radiance) of a rough surface with finite transverse dimensions comprises a "diffraction" lobe centered around the specular direction, along with numerous sidelobes known as speckle grains [32]. When studied in more detail, we will find that the central lobe resembles the diffraction of light by a smoothed-out version of the finite-size surface, where the rough surface features are 'homogenized' and appear akin to a thin film. As the transverse dimension increases, the specular lobe narrows, and the speckle grains get smaller. In the asymptotic case of infinitely large surfaces, the specular lobe converges to a Dirac delta distribution in the specular direction, while the speckle patterns, marked by small-scale fluctuations, gradually transition into a broad diffuse lobe. This evolution occurs because, with surface expansion, an increasing number of randomly distributed scatterers contribute to the scattered amplitude with random phases. This cumulative effect, in accordance with the law of large numbers, leads to a far-field scattered intensity that approaches the expected value.

The distinction between specular and diffuse components for the scattered light intensity emerges as a pivotal consideration within the context of wave scattering by rough surfaces. These two components have different physical origins, as we will see below, and have significant implications for optical effects – for instance, their relative weight plays a decisive role in determining the glossiness or matteness of a surface.

Thus, in this Section dedicated to fundamentals, we start by outlining some basic concepts on the statistical properties of light scattering by surfaces. We then familiarize the reader with the Bidirectional Scattering Distribution Function (BSDF), a highly multidimensional radiometric function describing how disordered metasurfaces scatter light.

### 3.1 Specular and diffuse components of the scattered light

The distinction between diffuse and specular light can be understood from the statistical properties of the scattered light. In this framework, one considers separately the average and fluctuating values of physical parameters with the averages computed among a large statistical ensemble of independent realizations. The quantities of main interest are electromagnetic field characteristics, including the field itself, its intensity or root mean square, along with auxiliary quantities such as the Poynting vector and the energy density that a wave carries.

Formally, the electric field vector $\mathbf{E}$ produced by a disordered metasurface upon illumination is the sum of an incident field $\mathbf{E}_b$, which propagates in the background medium (homogeneous or layered) without the metaatoms, and a scattered field vector $\mathbf{E}_s$ created by the excited metaatoms, that is $\mathbf{E} = \mathbf{E}_b + \mathbf{E}_s$. Defining $\langle ... \rangle$ as the average over disorder realizations, one can further write the scattered field as the sum of its average and a fluctuating term that varies from realization to realization,

$$\mathbf{E}_s = \langle \mathbf{E}_s \rangle + \delta \mathbf{E}_s, \tag{1}$$

with $\langle \delta \mathbf{E}_s \rangle = \mathbf{0}$ by definition.

Figure 2 illustrates this decomposition using fully-vectorial simulations conducted on a finite metasurface composed of parallel infinitely long silicon cylinders in air. The metasurface is illuminated at a 30° angle of incidence by a planewave polarized parallel to the cylinder $y$-axis so that we may denote by $E_s \equiv \mathbf{E}_s \cdot \hat{\mathbf{y}}$ the scattered electric field.

The top panel shows the $y$-component of the scattered field produced by a specific disorder realization, thereby exhibiting a complex pattern in space. In contrast, the average field $\langle E_s \rangle$ in the central panel readily takes predominantly the form of a reflected light beam, whose principal direction is indicated by a black arrow. This average field describes the diffracted light from a homogenized metasurface characterized by effective parameters that vary non-uniformly across the $x$ direction due to boundary effects (Subsection 4.3). Could the simulations have been made on a statistically translationally-invariant (i.e., infinite) metasurface, the average field would have strictly been a reflected planewave. The fluctuating field $\delta E_s$ shown in the bottom panel constitutes the remaining speckle pattern and averages to zero (the quantity $\langle \delta E_s \rangle_X$ in the subpanels describe the fluctuating field averaged over $X$ disorder realizations).

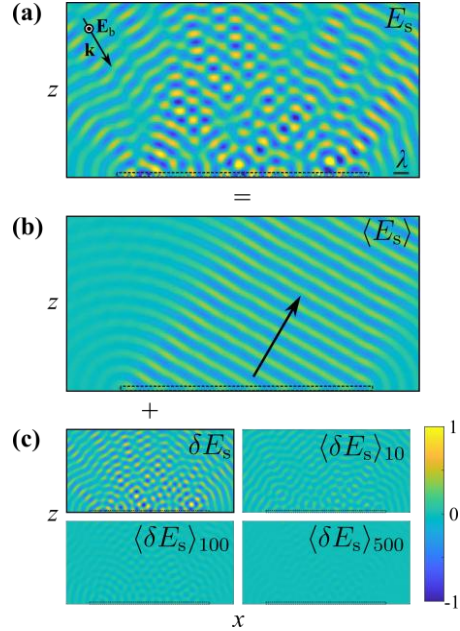

**Figure 2.** Illustration of the average (coherent) and fluctuating (incoherent) components of Equation 1 via fully-vectorial simulations for a finite-size disordered metasurface composed of identical silicon cylinders in air. The incident (background) field $\mathbf{E}_b$ is a planewave ($\lambda = 440$ nm) polarized along the $y$-direction and incident at a 30° angle relative to the surface normal. The cylinders, aligned parallel to the $y$-axis, are positioned using a random sequential addition algorithm with a non-overlapping condition. 2D simulations are conducted for assemblies of 15 infinitely long cylinders (140 nm diameter, $n_{Si} = 4.80 + 0.11i$) at a surface coverage of 30%, resulting in a side length of 7.0 μm ($\approx 16\lambda$). **(a)** Electric field, $E_s \equiv \mathbf{E}_s \cdot \hat{\mathbf{y}}$, scattered for a specific disorder realization. **(b)** The averaged field $\langle E_s \rangle$ is computed by averaging over $X = 1000$ independent disorder realizations. **(c)** $\langle \delta E_s \rangle_X$ denotes the fluctuating field averaged over $X$ realizations. In all plots, $E_s$ is normalized to the amplitude of the background field $|\mathbf{E}_b|$. Only the real parts of the complex fields are represented.

Now, let us examine the light intensity. Given $\langle \delta \mathbf{E}_s \rangle = \mathbf{0}$, we can readily demonstrate that the averaged intensity $\langle |\mathbf{E}_s|^2 \rangle$ of the light scattered by the metasurface can be decomposed into two components,

$$\langle |\mathbf{E}_s|^2 \rangle = |\langle \mathbf{E}_s \rangle|^2 + \langle |\delta \mathbf{E}_s|^2 \rangle. \tag{2}$$

The first term, $|\langle \mathbf{E}_s \rangle|^2$, previously qualified as the 'diffraction' lobe, corresponds to the specular component of the scattered light, whereas the second, $\langle |\delta \mathbf{E}_s|^2 \rangle$, corresponds to the diffuse component (the averaged speckle intensity). Note also that this decomposition also holds on the average total intensity, $\langle |\mathbf{E}|^2 \rangle = |\langle \mathbf{E} \rangle|^2 + \langle |\delta \mathbf{E}_s|^2 \rangle$.

The decomposition in Equation 2 is visually depicted in Figure 3. As anticipated, the specular component (central panel) primarily consists of a directional beam (white arrow). However, it is evident that this specular beam is significantly affected by numerous fringes, a consequence of finite-size effects that would vanish with simulations conducted on an infinite metasurface. As will be explained in Subsection 4.3, boundary effects are not confined to the metasurface boundaries but spread over the entire metasurface area. Quite in contrast, the diffuse component (bottom panel) exhibits a more uniform spatial distribution, resembling a broad diffuse lobe covering the entire hemisphere in reflection (and similarly in transmission, although not shown), owing to the subwavelength dimensions of the cylinders.

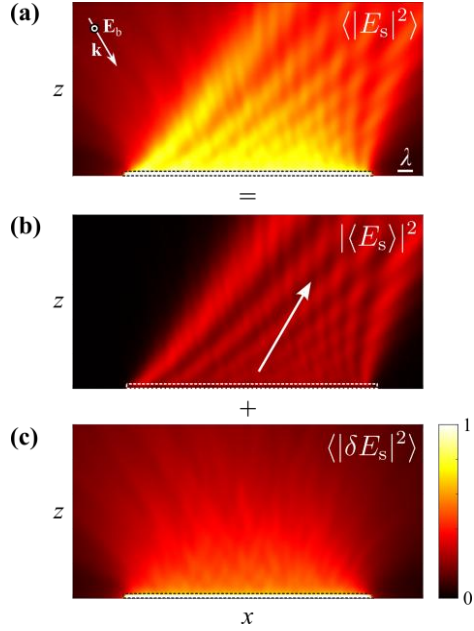

**Figure 3.** Illustration of the decomposition of the average scattered intensity into specular and diffuse components (Equation 2). The parameters used in the simulations are the same as those in Figure 2.

### 3.2 Bidirectional scattering distribution function

Every surface inherently possesses a limited transverse dimension. Nonetheless, when analyzing the refraction of light at an interface, it is a common and useful practice to treat interfaces as if they were infinite. This approach allows us to define the renowned Fresnel coefficients, which are fundamental parameters of significant importance in optics.

Similarly, for statistically translationally invariant metasurfaces, it is possible to define scattering coefficients that describe how metasurfaces scatter light for all possible planewave illuminations. The coefficients are also fundamental quantities. They were introduced in the 1960s [33,34] and are known in radiometry of random rough surfaces as the BSDF, a quantity that encompasses the BRDF for the Reflectance and the BTDF for the Transmittance.

Physically, the BSDF describes how, *on average*, a surface that is statistically invariant by translation scatters light for all possible planewave illuminations. Formally, the BSDF $f_s$ relates an incoming irradiance $E_i$ (in $W \cdot m^{-2}$) to a scattered radiance $L_s$ (in $W \cdot m^{-2} \cdot sr^{-1}$),

$$f_s(\hat{\mathbf{k}}_s, \hat{\mathbf{e}}_s, \hat{\mathbf{k}}_i, \hat{\mathbf{e}}_i, \omega) = \frac{<dL_s(\hat{\mathbf{k}}_s, \hat{\mathbf{e}}_s, \omega)>}{dE_i(\hat{\mathbf{k}}_i, \hat{\mathbf{e}}_i, \omega)}, \tag{3}$$

where $\omega$ is the frequency, $\hat{\mathbf{k}}_i$ and $\hat{\mathbf{k}}_s$ are the (unit) wavevectors of the incident and scattered planewaves, and $\hat{\mathbf{e}}_i$ and $\hat{\mathbf{e}}_s$ are their (unit) polarization vectors, respectively. Note the differentiated quantity on the right-hand side, which is used to reflect the fact that a measured power should necessarily involve an integral in the wavevector space [35]. Note also that the statistical average (here, performed on the positions of the particles and denoted with angular brakets) is essential in the definition of the BSDF to reconcile the coherent nature of wave scattering with the radiative transfer framework [36].

Following Subsection 3.1, it is convenient to decompose the BSDF for disordered metasurfaces into a sum of a specular and a diffuse term as $f_s = f_{s,\text{spe}} + f_{s,\text{dif}}$. The ratio between these two components is relevant in many practical situations and applications, constituting, for instance, a quantitative measure of the glossiness or matteness of a surface. Perfectly diffuse (matte) surfaces have $f_{s,\text{spe}} = 0$.

## 4. Electromagnetic analysis of finite-size disordered metasurfaces

Modeling the interaction of electromagnetic waves with disordered media is a notoriously challenging problem [17,37-38]. When the disordered medium consists of inhomogeneities much larger than the wavelength, traditional diffraction theory based on scalar optics (such as Fresnel or Fraunhofer diffraction) provides a useful framework for modeling. However, when the inhomogeneities are comparable to or smaller than the wavelength, as in the case of metasurfaces with resonant metaatoms, the interaction requires solving Maxwell's equations to account for electromagnetic diffraction [39]. This makes the analysis significantly more complex.

To simplify the analytical treatment, many theoretical studies assume the scatterers to behave as (point-like) electric dipoles in a homogeneous background medium and sometimes even the electric field to be a scalar quantity [40,41]. The problem is even more challenging for metasurfaces because it combines coherent effects at the subwavelength scale and multiple scattering at the mesoscale: resonant multipolar particles cannot be modeled as Rayleigh scatterers; the presence of a substrate changes the resonance frequencies of the nanoparticles through mode hybridization; and the monolayer geometry has a strong anisotropy, not encountered in 3D complex media.

We thus anticipate that the prediction of the BSDF for statistically translationally invariant metasurfaces containing resonant particles on layered substrates remains a significant challenge [17]. Two complementary strategies are available to address this challenge. The first approach involves solving Maxwell equations with utmost precision (this Section) for progressively larger metasurfaces, ultimately allowing us to extrapolate the BSDF for infinitely large metasurfaces. The second approach (Section 5) relies on approximate models.

### 4.1 Theoretical framework

Let us start by formally defining the electromagnetic problem of interest. We consider a collection of $N$ particles located at coordinates $\mathbf{r}_j = [x_j, y_j, z_j]$ with $j = 1 \ldots N$. These particles exhibit variations in composition, dimensions, and morphology. They may be dispersed within a uniform medium or a layered substrate – with Oz the axis of the Cartesian coordinate system that is normal to the substrate. All the layers are assumed to be optically isotropic and nonmagnetic. They are thus characterized by a single scalar, their respective refractive index. The system is excited by an electromagnetic wave at frequency $\omega$. For the sake of generality, we introduce a variable $\mathbf{\Psi}$ to denote the electromagnetic fields. This could encompass the electric field, $\mathbf{E}$, or the magnetic field, $\mathbf{H}$, in Cartesian coordinates, as well as a representation of these fields in vector spherical wave functions, among other possibilities.

As defined above, the total field $\mathbf{\Psi}$ is the sum of the field $\mathbf{\Psi}_b$ propagating in the background medium without any particle and the field $\mathbf{\Psi}_s$ scattered by the assembly of particles,

$$\mathbf{\Psi} = \mathbf{\Psi}_b + \mathbf{\Psi}_s. \tag{4}$$

The background field $\mathbf{\Psi}_b$ is known in general, and the problem is to determine the scattered field $\mathbf{\Psi}_s$.

For discrete media composed of particles, the scattered field $\mathbf{\Psi}_s$ equals the sum of the field scattered by each individual particle $j$. The total field observed at a point of space (labeled as o) can then be expressed in general terms as

$$\mathbf{\Psi}^{(o)} = \mathbf{\Psi}_b^{(o)} + \sum_{j=1}^{N} \mathbf{P}_b^{(o,j)} \mathbf{T}^{(j)} \mathbf{\Psi}_{\text{exc}}^{(j)}. \tag{5}$$

In this expression, $\mathbf{\Psi}_{\text{exc}}^{(j)}$ is the field exciting the particle $j$, $\mathbf{T}^{(j)}$ is the transition operator associated to particle $j$, describing how the exciting field polarizes particle $j$, and $\mathbf{P}_{\text{b}}^{(o,j)}$ is an operator describing the propagation of the field radiated by the polarized particle $j$ to the observation point o in the background medium. $\mathbf{P}_{\text{b}}^{(o,j)}$ may be an integral operator using the dyadic Green's function, in which case $\mathbf{T}^{(j)}$ would be an integral operator as well, or a superposition of vector spherical wave functions, in which case $\mathbf{T}^{(j)}$ would be the renowned T-matrix.

The multiple scattering problem appears when considering that the exciting field on particle $j$ is not only the background field at the position of particle $j$, but also the field scattered by all other particles ($l \neq j$) on particle $j$,

$$\mathbf{\Psi}_{\text{exc}}^{(j)} = \mathbf{\Psi}_{\text{b}}^{(j)} + \sum_{\substack{l=1 \\ l \neq j}}^{N} \mathbf{P}_{\text{b}}^{(j,l)} \mathbf{T}^{(l)} \mathbf{\Psi}_{\text{exc}}^{(l)}, \tag{6}$$

where $\mathbf{P}_{\text{b}}^{(j,l)}$ now describes the propagation of the field from particle $l$ to particle $j$.

Numerically, the solution of Equation 6 for a finite number $N$ of particles involves matrix inversion, yielding the exciting field $\mathbf{\Psi}_{\text{exc}}^{(j)}$, which can subsequently be used in Equation 5 to predict the total scattered field $\mathbf{\Psi}^{(o)}$. For disordered media, reliable predictions for the average field and the average intensity typically demand achieving a statistical averaging over many disorder realizations. An inherent limitation of this approach is that only finite numbers of particles can be simulated in practice, a topic further elaborated in the following subsections.

Analytically, addressing the problem entails employing perturbative techniques for ensemble-averaged quantities, assuming statistically translationally-invariant media [37,41]. This involves deriving approximations for the average exciting field from Equation 6, the average total field from Equation 5, and the average total intensity. Nevertheless, the drawback of these analytical treatments lies in their inherent approximative nature. Solutions are often restricted to first or second-order expansions, disregarding near-field coupling effects between particles or neighboring interfaces, and apply primarily to media with reasonably low filling fractions.

## 4.2 Full-wave electromagnetic simulations

The electromagnetic modeling of large-area metasurfaces, whether ordered or disordered, demands significant computational resources [42-48]. Despite many similarities, ordered and disordered metasurfaces are rarely modeled with the same approaches.

**General Maxwell solvers.** Full-wave numerical approaches, which discretize the entire computational domain—including absorbing boundaries—are commonly employed for simulating ordered metasurfaces or two-phase disordered metasurfaces. Established Maxwell solvers are well-suited for this purpose, with the Rigorous Coupled Wave Analysis (RCWA) being a typical choice [49,50]. RCWA is particularly advantageous for modeling metasurfaces with binary profiles, where the fields within the binary structure are expanded into eigenmodes, often referred to as Bloch modes due to their periodic nature. A notable feature of RCWA is its analytical treatment of fields in the substrate and superstrate using plane wave expansions. This eliminates the need for Perfectly Matched Layers (PMLs) to enforce outgoing wave conditions in open half-spaces.

The computational bottleneck in RCWA arises from solving the eigenvalue problem associated with eigenmodes and the method quickly reaches practical limitations on the size of the unit cell that can be modeled. For example, while RCWA excels in accuracy for dielectric metasurfaces and is often the preferred choice, it struggles to accommodate supercells larger than $3\lambda \times 3\lambda$. This limitation makes RCWA suitable for certain studies but insufficient for modeling large-scale disordered metasurfaces.

Another highly effective technique that avoids the aforementioned bottleneck is the finite-difference time-domain (FDTD) method [51]. In FDTD, Maxwell's equations are discretized on a Yee grid in both space and time, resulting in computations that involve only basic arithmetic operations, primarily addition and subtraction. The convergence is rather slow compared to the RCWA and achieving high accuracy often requires a spatial discretization of around 20 points per wavelength within high-index dielectric materials. Despite this, FDTD's reliance on simple algebra makes it highly amenable to parallelization on high-performance computing (HPC) architectures using GPUs or CPUs. This scalability enables the simulation of large, disordered or ordered metasurfaces on powerful computing clusters. For instance, FDTD can efficiently compute the dispersive response of metasurfaces with unit cells as large as $\sim 20\lambda \times 20\lambda$ in a single simulation [52]. Although the finite element method (FEM) provides an advantage with adaptive spatial discretization tailored to the shape of scatterers—particularly beneficial for metallic structures—this precision comes at the cost of limited computational domains. Similar to RCWA, FEM generally restricts the transverse dimension to just a few wavelengths [53]. This limitation makes FDTD a more suitable choice for handling extensive metasurfaces.

Dedicated software tools, often utilizing parallelization techniques with GPU-based implementations, are actively being developed for the analysis of large-scale, ordered metasurfaces [53-57]. These tools, designed to handle complex computational demands, can also be effectively applied to the analysis of disordered metasurfaces, broadening their utility across various metasurface configurations.

**Dedicated Maxwell solvers for particulate metasurfaces.** In contrast, disordered particulate metasurfaces benefit significantly from specialized numerical techniques. The latter leverage the distinctive morphologies of these structures, increasing their efficiency by precomputing light scattering by the individual scatterers that compose the metasurface. Subsequently, an analytical treatment is applied to model the interaction between the scatterers and the outgoing wave conditions (Subsection 4.1).

If one assumes that $P$ degrees of freedom are required to accurately mesh the individual scatterers of a metasurface with $N$ scatterers, the solution is typically computed by inverting a matrix of size $(N \times P)^2$. We immediately see that, for a given $N$, the numerical efficiency lies in the possibility to pre-model the scatterer responses with a number $P$ of degrees of freedom as small as possible. Assessing $P$ is, however, a complex task, as it is contingent on the chosen methodology and necessitates the consideration of various factors, such as near-field coupling between neighboring scatterers, intricate shapes, and multipolar resonant reactions.

In the widespread T-matrix method, the field outside the scatterer is expanded using spherical vectorial wave functions and a T-matrix is used to describe how each scatterer transforms multipolar incident fields into scattered field ones. This approach, initially developed for acoustic waves [58], has been extensively explored over decades [59]. It is nowadays well mastered and multiple open-source implementations are available [60]. Since the matrix is not sparse, the computational time increases significantly for very large samples. Typically, light scattering by hundreds or thousands of scatterers can be self-consistently analyzed with standard personal computers. The great strength of the approach lies in its efficiency in modeling the far-field interaction with only a few spherical vectorial wave functions. However, because expansions with spherical vectorial wave functions are valid exclusively outside a circumscribing sphere that encompasses each scatterer (Figure 4), the T-matrix method has some intrinsic limitations: Firstly, the circumscribing spheres of two nearby scatterers should not overlap, which is restrictive for situations involving elongated scatterers at high densities. Secondly, these spheres must not intersect an interface [61]. Therefore, treating problems involving non-spherical scatterers deposited on a substrate necessitates more advanced expansions, as we will see.

These limitations have given rise to several theoretical works in recent years to alleviate their impact [62-66] at the expense of heightened complexity and occasionally a trade-off in stability. Nevertheless, impressive results have been achieved over time. Notably, the freely available software package SMUTHI [67], introduced recently, circumvents the interface challenge by coupling the T-matrix approach for individual particles with the S-matrix approach for thin-film stack computations, as proposed in [68].

The inherent limitation of the T-matrix can be circumvented by representing the scatterers as ensembles of Dirac sources and using dyadic Green's functions in the background medium to describe their interaction. In the popular Discrete Dipole

Approximation, each particle is discretized into a dense array of small polarizable elements that, according to effective medium theory, behave collectively as the real material composing the particle. The Discrete Dipole Approximation is relatively simple, intuitive, and physically transparent. However, many dipoles are often needed per particle, and only small clusters of particles (typically around 10 [69]) can be accurately analyzed in practice.

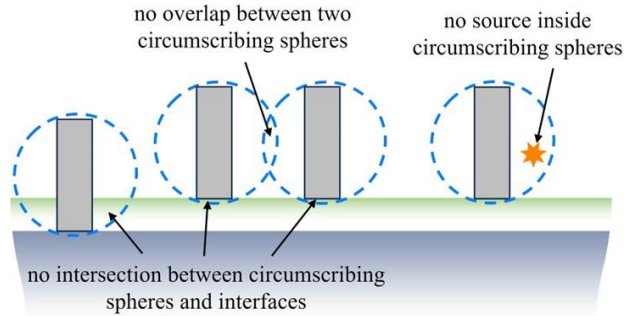

**Figure 4** Limitations of the T-matrix approach. The blue dashed circles represent the circumscribing spheres around each particle (illustrated as rectangles). For the approach to remain valid, these spheres must not overlap with one another or with interfaces, and the driving source must be positioned outside all the spheres.

In the more advanced surface integral equation method, particles are replaced by equivalent electric and magnetic currents distributed over their boundary surfaces (via Love's equivalence principle) [70]. Thus, only the surfaces are "discretized" and, compared to the Discrete Dipole Approximation, a high accuracy is achieved with much smaller $P$'s. This approach, inherited from the radio wave community, considerably lowers computational loads, and impressive results have been obtained for large and dense ensembles with $N \approx 1000$ particles and complicated metallic scatterers, e.g., nanostars, on high-performance parallel computers [71]. The only singular constraint lies in the current implementation, which exclusively accommodates free-space Green's functions and may thus only analyze layered substrates possessing finite transversal and longitudinal dimensions. But this is not a fundamental limitation [72].

In the more recent global polarizability matrix method [73], small $P$ values are attained by substituting the scatterers with a restricted set of numerical dipoles. Unlike the physical dipoles implemented in the DDA, numerical dipoles respond directly to the incident field and to the fields that illuminate the other dipoles. Thus, even though only $P$ dipoles are incorporated per particle, there are $P^2$ hidden degrees of freedom accessible for precomputing the spatially nonlocal polarizability matrix of the dipole set, named the global polarizability matrix, which describes how each individual scatterer interacts with the electromagnetic fields. The approach, therefore, shifts the entire burden of the electromagnetic problem to the precomputation of the polarizability matrix method by resolving an inverse problem. This step is more demanding than a T-matrix computation, but for small $P$ values of ~50, metasurfaces with several hundred scatterers can be studied with desktop computers, including non-spherical particles in close vicinity to each other or to interfaces, such as in Figure 4.

### 4.3 Extrapolation towards infinite-size metasurfaces

With access to robust simulation tools capable of analyzing spatially extended, albeit finite, metasurfaces, let us now consider how these tools may be used to deduce the properties of laterally infinite metasurfaces. It is important to bear in mind that this "extrapolation" step is challenging, primarily due to substantial boundary errors.

To illustrate this challenge, let us examine a scenario involving a circular metasurface comprised of "Huygens" nanospheres with identical magnetic and electric dipolar polarizabilities. We compute the mean field just above or below the metasurface on the $z$- axis, specifically the further away from the circular metasurface boundary (Figure 5a). As the metasurface radius, denoted as $R$, increases, one would naturally anticipate that the average field value starts to approach the mean field response of the infinite metasurface. However, it is evident that artifacts arising from conducting simulations for finite-size domains persist, even when working with relatively large domains, making the achievement of convergence a challenging endeavor (Figure 5b).

The slow oscillatory convergence can be explained by observing that the number of particles in a ring of radius $R$ and thickness $dR \ll R$ (Figure 5a) is proportional to $R$ while the field created in the center of the disc by any of these particles has an amplitude of the order of $\exp(ikR)/R$. Thus, similarly to the field created at the center of a charged disc of uniform charge density, the field generated by an extra layer of particles upon increasing $R$ is not expected to decrease with $R$. Additionally, neglecting polarization issues, all particles within a ring add an identical phase to the field in the center of the system. Thus, rings with a radius that is an exact multiple of the wavelength make a positive contribution, while those with an additional half-wavelength in radius contribute negatively. This intuitive explanation accounts for the gradual and oscillatory convergence observed as $R$ increases.

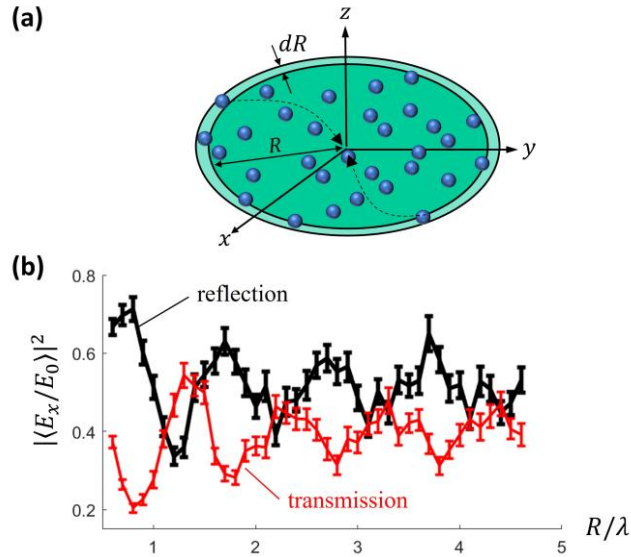

**Figure 5** Intrinsic difficulty in computing the BSDF. **(a)** A monolayer of Si nanospheres (80nm radius, 10 μm$^{-2}$ density) in air, encased in a disc of radius $R$, is illuminated by a normally incident monochromatic planewave ($\lambda = 520$ nm) polarized parallel to the $x$-axis. **(b)** $R$-dependence of the average electric field $|\langle E_x \rangle|^2$ computed just below ($z = -0.35\lambda$, red) or above ($z = +0.35\lambda$, black) the monolayer. The numerical data are averaged over 100 independent disorder realizations. The nanosphere centres are distributed in the disc using a random sequential addition algorithm with a 100 nm-radius exclusion disc, corresponding to a packing fraction $p = 0.3$. The electromagnetic computations are performed assuming that the nanospheres exhibit dipolar electric and magnetic polarizabilities.

These finite-size artifacts make it difficult to infer the characteristic properties of infinite metasurfaces based on electromagnetic computations conducted for finite metasurfaces and have prompted the development of dedicated methods to mitigate their impact. Two primary approaches have emerged to address this issue.

In the field-stitching approach, Maxwell equations are solved for numerous small square subdomains, each corresponding to an independent realization. Subsequently, these subdomains are combined to artificially expand the metasurface area, and the fields at the boundaries of these subdomains are smoothed to prevent undesired field discontinuities. This method has found applications in various electromagnetic problems, such as predicting the behavior of diffractive lenses, metalenses [74-76], or antenna arrays [77], especially when dealing with large-scale lenses that cannot be directly computed. However, its application to random surfaces is less common. Notably, a recent study [78] demonstrated compelling results, showcasing reasonable agreement between predictions and optical measurements for the absorbance and reflectance of silicon surfaces coated with random pyramids to reduce reflection. Despite these successes, the field stitching method remains somewhat empirical, lacking a robust mathematical foundation and making the accuracy of its results challenging to anticipate.

In the supercell approach, identical subdomains are seamlessly connected using pseudo-periodic conditions. This artificial periodicity effectively mitigates the boundary effects commonly encountered in finite-size computations. Generally, it yields satisfactory results, as demonstrated in [79], where stabilized numerical results are presented for both normal and oblique incidences in specular and diffuse reflections as the periodicity is increased. However, there are many exceptions to this rule. In a recent investigation of metamaterial homogenization [80], researchers observed unphysical oscillations in the specular transmission spectra of metamaterials as the wavelength varied. They suggest introducing a linear variation in the supercell period with wavelength to reduce oscillations.

While the findings in [79,80] are indeed promising, it is important to acknowledge that the reliability of the supercell approach is uncertain. Like the stitching method, the approach is built upon the assumption that global couplings can be safely disregarded. Indeed, it is anticipated that the impact of long-range electromagnetic interactions (Figure 5) is diminished in disordered media due to positional randomness, but the specific conditions under which this argument holds true remain unclear with an artificial periodization.

The field-stitching and supercell approaches provide reasonably accurate results, with relative errors typically below ten percent. Nevertheless, both methods demand a substantial number of computations to evaluate statistical convergence, and there is limited guidance on predicting when convergence will be rapid or slow. Furthermore, these methods offer limited insights into the underlying physics. Additionally, up to this point, most investigations have primarily focused on the specular component of the BSDF within the context of homogenization studies, with minimal to no available results for the diffuse component. The challenge of effectively simulating laterally infinite metasurfaces persists, emphasizing the necessity for approximate models.

## 5.  Approximate analysis of disordered metasurfaces with infinite lateral dimensions

Classical theories on wave scattering by rough surfaces [1] often focus on two extremes: surfaces with imperfections significantly smaller than the wavelength, as described by the Rayleigh-Rice theory, or surfaces that are locally flat but randomly angled, with features significantly larger than the wavelength, such as those modeled by the tangent-plane Kirchhoff approximation. However, these frameworks are inadequate for analyzing light scattering by disordered metasurfaces comprising well-defined particles (or inclusions) on layered substrates. For an accurate calculation of the BSDF, it is necessary to consider multiscale effects, such as the multipolar resonances of individual particles, the hybridization of the resonances with their mirror images in the substrate [81-85], and the mutual interaction between particles [81,86]. Due to the complex nature of these interactions, the existing literature offers only a few models addressing this complexity.

An essential aspect of BSDF models is their ability to provide a deeper insight into the optical phenomena involved and help design studies. Hence, this Section significantly emphasizes the Independent Scattering Approximation (ISA). This approximation essentially assumes that particles scatter light independently from the interaction with other particles. Consequently, the field driving a given particle, $j$, is solely the background field incident on the particle, i.e., $\boldsymbol{\Psi}_{\text{exc}}^{(j)} \approx \boldsymbol{\Psi}_{\text{b}}^{(j)}$ in Equation 6. It is evident that this constitutes a rudimentary approximation, confining its applicability to highly dilute systems. Despite this inherent limitation, the ISA carries substantial significance: it elucidates fundamental principles in a remarkably straightforward manner and serves as the foundation for more sophisticated theories, as we will underscore in Subsection 5.1.

Thus, we start in Subsection 5.1 by demonstrating how specific approximations enable the explicit separation of contributions from individual metaatoms and their spatial arrangement. In Subsection 5.2, we derive comprehensive expressions for the specular and diffuse components of the intensity scattered by finite-size disordered metasurfaces and introduce the structure factor—an essential parameter in scattering theory. Moving on to Subsection 5.3, we provide analytical expressions for the specular and diffuse components of the BSDF for infinite-size disordered metasurfaces under the ISA. Finally, in Subsection 5.4, we review more advanced BSDF models that address certain limitations of the ISA and discuss the remaining challenges in modeling.

## 5.1 Individual versus collective response

We consider a metasurface composed of a collection of $N$ identical particles within a multilayered substrate, positioned $\mathbf{r}_j = [\mathbf{r}_{j,\parallel}, z_j = 0]$ with $j = 1 \ldots N$, with $Oz$ representing the axis of the Cartesian coordinate system perpendicular to the substrate. The metasurface is illuminated by a planewave at frequency $\omega$. The background electric field $\mathbf{E}_b$ is a superposition of two planewaves with equal in-plane wavevectors. For simplicity, we assume the particles are situated on the substrate, although the same principles apply to embedded particles.

The background field can be expressed as $\mathbf{E}_b(\mathbf{r}) = E_i \hat{\mathbf{e}}_i \exp[i\mathbf{k}_i \cdot \mathbf{r}] + r_{sub} E_i \hat{\mathbf{e}}_r \exp[i\mathbf{k}_r \cdot \mathbf{r}]$, where $E_i$ is the amplitude of the incident planewave, $r_{sub}$ is the reflection coefficient of the substrate, $\hat{\mathbf{e}}_i$ and $\hat{\mathbf{e}}_r$ are the polarization directions of the incident and reflected planewaves, and $\mathbf{k}_i = [\mathbf{k}_{i,\parallel}, k_{i,z}]$ and $\mathbf{k}_r = [\mathbf{k}_{i,\parallel}, -k_{i,z}]$ are their wavevectors. There are two possible orthogonal polarizations, i.e. $\hat{\mathbf{e}}_i$ and $\hat{\mathbf{e}}_r$ corresponds to either TE or TM polarized planewaves. Within the ISA, each particle of the metasurface is excited by the same field up to a phase factor $\exp[i\mathbf{k}_{i,\parallel} \cdot \mathbf{r}_{j,\parallel}]$. Similarly, due to reciprocity, each particle radiates the same scattered field up to a phase factor $\exp[-i\mathbf{k}_{s,\parallel} \cdot \mathbf{r}_{j,\parallel}]$. Thus, the field scattered by the collection of identical particles can be written as the product of the field scattered by an isolated particle $\mathbf{E}_s^{(0)}$ and a dephasing term summed over all particles,

$$\mathbf{E}_s(\mathbf{r}) = \mathbf{E}_s^{(0)}(\mathbf{r}) \times \sum_{j=1}^{N} \exp[-i\mathbf{q}_\parallel \cdot \mathbf{r}_{j,\parallel}], \tag{7}$$

where we have introduced the in-plane scattering wavevector $\mathbf{q}_\parallel = \mathbf{k}_{s,\parallel} - \mathbf{k}_{i,\parallel}$.

The first term on the right-hand side of Equation 7, $\mathbf{E}_s^{(0)}$, represents the field scattered by an individual particle positioned at $\mathbf{r}_\parallel = [0,0]^T$. It depends only on the particle composition, shape, size, the nature of the layered environment, and the frequency, polarization, and direction of the incoming planewave. When our interest lies primarily in the BSDF, attention can be limited to the (far-field) scattering amplitudes of the particle. These amplitudes indicate how the scattered planewaves are distributed angularly (Figure 6b). Typically, they are calculated using a Maxwell equations solver, followed by a near-to-far-field transformation [87], and are related to the form factor that is precisely defined in Appendix C.

The second term in Equation 7, $\sum_{j=1}^{N} \exp[-i\mathbf{q}_\parallel \cdot \mathbf{r}_{j,\parallel}]$, describes the dephasing between the incident and scattered planewaves due to the particle position relative to the origin. It typically becomes negligible under statistical averaging if $\mathbf{q}_\parallel \neq \mathbf{0}$, as illustrated in Figure 2 for the averaged speckle field. However, when focusing on speckle intensity, this term, influenced by $\mathbf{q}_\parallel$, can introduce notable spectral and angular characteristics in the scattered intensity. Hence, it represents a significant parameter to consider in design studies.

In the following Subsection, we will utilize the decomposition outlined in Equation 7 to formally derive two crucial variables that influence the scattering behavior of periodic or disordered arrays of particles: the form and structure factors. These factors offer valuable insights for harnessing the BSDF. Hence, it is important to assess the scope within which this decomposition remains valid.

If any assumption underlying Equation 7 is invalid, the description no longer applies. There are three scenarios to mention specifically. Firstly, in collections of polydisperse particles; secondly, in cases of non-uniform illuminations, such as Gaussian-beam illumination, where the background field varies noticeably with the lateral position of the particle; thirdly, when particles are situated within a layered environment and are not constrained to a single plane ($z_j \neq$ cst), leading to variations in both the background field and the scattered field $\mathbf{E}_s^{(0)}(\mathbf{r})$ with the vertical position.

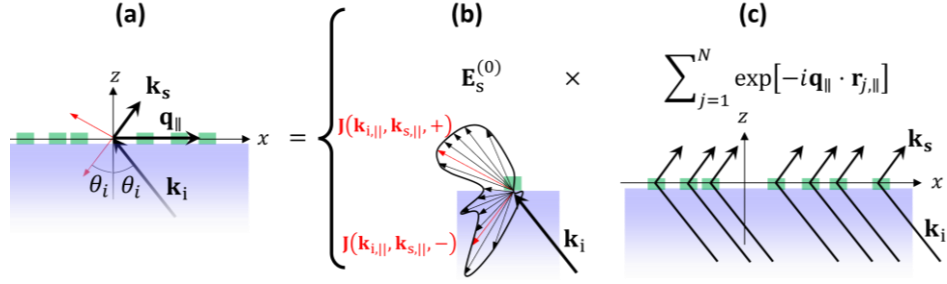

**Figure 6** The ISA (or mean-field approach) transforms a complex many-body interaction scenario **(a)** into two much simpler components: an effective one-body problem **(b)** and a multiple-beam interference phenomenon **(c)**. **(a)** A monolayer of identical particles is illuminated by a planewave, with $\mathbf{q}_\parallel = \mathbf{k}_{s,\parallel} - \mathbf{k}_{i,\parallel}$ representing the difference between the in-plane (parallel) wavevectors of the scattered ($\mathbf{k}_s = [\mathbf{k}_{s,\parallel}, k_{s,z}]$) and incident ($\mathbf{k}_i = [\mathbf{k}_{i,\parallel}, k_{i,z}]$) planewaves. The red arrows indicate the specular directions of the reflected and transmitted waves. **(b)** The one-body problem is characterized by two infinite sets of $2 \times 2$ Jones matrices, $\mathbf{J}(\mathbf{k}_{i,\parallel}, \mathbf{k}_{s,\parallel}, +)$ and $\mathbf{J}(\mathbf{k}_{i,\parallel}, \mathbf{k}_{s,\parallel}, -)$, which provides the linear relationship between the amplitudes of the incident planewave and all the planewaves scattered by a single particle positioned at the origin ($\mathbf{r}_\parallel = [0,0]^T$). The differential scattering cross section in $m^2 \cdot sr^{-1}$, also known as the form factor, is simply $\|\mathbf{J}(\mathbf{k}_{i,\parallel}, \mathbf{k}_{s,\parallel}, \pm)\|^2$, see Equation (9) and Equation B7 in Appendix B. Appendix C describes how to compute the Jones matrices. The two red arrows highlight the amplitudes, $\mathbf{J}(\mathbf{k}_{i,\parallel}, \mathbf{k}_{i,\parallel}, \pm)$, scattered in the two specular directions; they are involved in the specular contribution to the BSDF, see Equations 15a and 15b. **(c)** The many-body system manifests as a multi-beam interference, contributing to the structure factor $S(\mathbf{q}_\parallel)$, which plays a key role in the diffuse contribution to the BSDF, see Equations B8 or B12 in Appendix B.

While the ISA may appear rudimentary to readers, it is crucial to note that a higher-order mean-field approximation yields the same decomposition. Consider a scenario where the field exciting particle $j$ includes not only the background field arriving on particle $j$ but also the sum of the background field and the average scattered field generated by all other particles $l \neq j$, $\mathbf{E}_{exc}^{(j)} \approx \mathbf{E}_b^{(j)} + \langle \mathbf{E}_s^{(j)} \rangle$. Here, the electromagnetic interaction is considered in a mean-field sense, with only the field fluctuations due to multiple scattering being neglected. For metasurfaces that are statistically invariant under translation (i.e., of infinite size) under planewave illumination, the average scattered field also comprises a superposition of two planewaves (forward and backward propagating) with equal in-plane wavevectors. Consequently, each particle within a metasurface is excited by the same field and emits the same scattered field, leading to Equation 7. Importantly, the validity of the decomposition in Equation 7 beyond the ISA provides an intuitive explanation for the resilience of the concept of the structure factor, which will be discussed in the subsequent Subsection. This renders it applicable even for moderately high particle densities.

In this mean-field approach, each individual particle is "dressed" by the other particles, resulting in a modification of its scattering properties [via $\mathbf{E}_s^{(0)}$ in Equation 7], by introducing an "effective polarizability" for instance [88]. It is worth noting that, to predict sophisticated phenomena such as Anderson localization or enhanced backscattering, more advanced approximations beyond the framework of the structure factor concept are required [38,17,89].

Finally, it is important to acknowledge that the initial stages in the design of ordered metasurfaces, such as metalenses, rely on a similar level of approximation. The generation of lookup tables often assumes a periodic arrangement and thus imposes a specific electromagnetic interaction among the meta-atoms. This introduces a systematic error into the design process, similar to that inherent to the mean-field approximation in disordered systems [90].

### 5.2 Form and structure factors

In this Subsection, we derive a closed-form expression for the average scattered intensity using the ISA. The approach allows us to demonstrate how the fundamental notions of form and structure factors, originating from crystallography, influence the scattering characteristics of metasurfaces.

Consistently with Section 3, statistical averaging over disorder realizations is denoted by $\langle ... \rangle$. From Equation 7, the average scattered intensity by monodisperse metasurfaces is given by

$$\langle |\mathbf{E}_s(\mathbf{r})|^2 \rangle = |\mathbf{E}_s^{(0)}(\mathbf{r})|^2 \langle \sum_{l=1}^N \sum_{j=1}^N \exp[i\mathbf{q}_\parallel \cdot (\mathbf{r}_{l,\parallel} - \mathbf{r}_{j,\parallel})] \rangle. \tag{8}$$

The first term on the right-hand side is the intensity scattered by a (fictitious) single particle positioned at the origin ($\mathbf{r}_\parallel = [0,0]^T$) of the Cartesian coordinate in the presence of the layered substrate. It promotes the field $\mathbf{E}_s^{(0)}$ as a key actor of our analysis. In Appendix A, we introduce the radiation diagram of the electromagnetic field scattered by the fictitious particle by performing a planewave expansion of $\mathbf{E}_s^{(0)}$. The expansion requires the knowledge of two infinite sets (respectively labeled '+' and '−' for the superstrate and substrate) of dyadic $2 \times 2$ tensors. These tensors, usually called Jones matrices, $\mathbf{J}(\mathbf{k}_{i,\parallel}, \mathbf{k}_{s,\parallel}, \pm)$, contain the scattering coefficients between the incident and scattered planewaves in the basis formed by their two possible orthogonal polarizations (e.g., TE and TM), see Figure 6 and Figure 13 in Appendix A.

The *form factor*, a quantity often used in crystallography and known as the scattering diagram in radio science, is defined as the 'differential scattering cross section' in $\text{m}^2 \cdot \text{sr}^{-1}$ of the fictitious particle (Figure 6b). For an incident planewave with a polarization parallel to the vector $\hat{\mathbf{e}}_i$, it is given by

$$F(\mathbf{k}_{s,\parallel}, \pm, \mathbf{k}_{i,\parallel}, \hat{\mathbf{e}}_i) = |\mathbf{J}(\mathbf{k}_{i,\parallel}, \mathbf{k}_{s,\parallel}, \pm) \hat{\mathbf{e}}_i|^2. \tag{9}$$

The Jones matrices will play a key role in the following. Appendix C explains how they can be computed using near-to-far-field transformations.

Back to Equation (8), note the double sum over particles $j$ and $l$ in the second term, which can be conveniently written as

$$\langle |\mathbf{E}_s(\mathbf{r})|^2 \rangle = N |\mathbf{E}_s^{(0)}(\mathbf{r})|^2 S(\mathbf{q}_\parallel), \tag{10}$$

where we introduced the *static structure factor*

$$S(\mathbf{q}_\parallel) = \frac{1}{N} \langle \sum_{l=1}^N \sum_{j=1}^N \exp[i\mathbf{q}_\parallel \cdot (\mathbf{r}_{l,\parallel} - \mathbf{r}_{j,\parallel})] \rangle, \tag{11}$$

a real-valued scalar quantity that expresses an interference effect between pairs of particles.

As shown in Subsection 3.1, the average scattered intensity $\langle |\mathbf{E}_s|^2 \rangle$ can be decomposed into a specular contribution $|\langle \mathbf{E}_s \rangle|^2$ and a diffuse contribution $\langle |\delta E_s|^2 \rangle$. The specular contribution is directly inferred from Equation 7

$$|\langle \mathbf{E}_s(\mathbf{r}) \rangle|^2 = |\mathbf{E}_s^{(0)}(\mathbf{r})|^2 |\langle \sum_{j=1}^N \exp[-i\mathbf{q}_\parallel \cdot \mathbf{r}_{j,\parallel}] \rangle|^2. \tag{12}$$

The second term on the right-hand side gives rise to diffraction lobes around the specular directions in both reflection and transmission. This can be understood for spatially uniform distributions, where the average summation in the second term corresponds to the Fourier transform of the finite-sized surface. The diffuse contribution is then obtained by subtracting Equation 12 from Equation 10

$$\langle |\delta E_s(\mathbf{r})|^2 \rangle = \langle |\mathbf{E}_s(\mathbf{r})|^2 \rangle - |\langle \mathbf{E}_s(\mathbf{r}) \rangle|^2 = N |\mathbf{E}_s^{(0)}(\mathbf{r})|^2 S_r(\mathbf{q}_\parallel). \tag{13}$$

Here, $S_r(\mathbf{q}_\parallel)$ denotes a *structure factor* tailored to exclusively contribute to the diffuse component. It is defined by

$$S_r(\mathbf{q}_\parallel) = S(\mathbf{q}_\parallel) - \frac{1}{N}\left|\left\langle \sum_{j=1}^{N} \exp[-i\mathbf{q}_\parallel \cdot \mathbf{r}_{j,\parallel}]\right\rangle\right|^2. \tag{14}$$

Equations 12 and 13, derived here within the ISA framework, describe the specular and diffuse components of the average scattered intensity. They highlight the significance of both the scattering characteristics of individual particles and their spatial arrangement in determining the total scattered intensity from disordered metasurfaces. These equations are specifically designed for finite-sized metasurfaces. In Subsection 5.3, we will present analytical formulations for the BSDF of infinitely large disordered metasurfaces. Practical examples of structure factors are given in Appendix D.

The structure factor $S_r(\mathbf{q}_\parallel)$ can be readily evaluated through statistical averaging for finite-size metasurfaces. Additionally, inverse methods exist, enabling the calculation of particle positions to achieve a desired factor. Various freeware programs or techniques [91-94] are available for generating complex structure factors, even those lacking radial symmetry [95]. Conversely, the precise determination of the scattering pattern requires solving Maxwell equations numerous times for different wavelengths, incident angles, and polarizations. In each case, a near-to-far-field transformation is performed to compute the radiation pattern (refer to Figure 6b).

### *5.3 BSDF model of infinite metasurfaces within the ISA*

As discussed in Section 4, accurately predicting the BSDF of infinite metasurfaces through full-wave simulations of Maxwell equations presents numerous challenges. Hence, approximate models become crucial despite their inherent reduction in precision.

In this Subsection, we elaborate on the diffuse and specular contributions to the BSDF of infinite metasurfaces within the ISA. We provide a concise overview of the main findings, deferring technical details of the derivation to Appendix A and B. Appendix A focuses on the specular component of the scattered light, while Appendix B deals with the diffuse component. Both Appendices begin by calculating the field scattered by a collection of $N$ particles. Subsequently, an averaging process over disorder realizations (i.e., statistical positions of particles) is performed. For this averaging, we assume that the metasurface is statistically translational invariant, with all particles independently positioned according to identical statistical distributions. Finally, we take the limit of an infinite metasurface ($N \to \infty$) while maintaining a fixed particle surface density $\rho$.

Although the appendices may appear technically intricate for newcomers, the results themselves are remarkably straightforward. This simplicity and transparency stem from the decomposition outlined in Equation 7, which transforms a many-body interaction problem into a simple one-body problem relying on the field $\mathbf{E}_s^{(0)}$ scattered by the fictitious particle positioned at the origin ($\mathbf{r}_\parallel = [0,0]^T$) of the Cartesian coordinate.

### Specular component of the scattered light

An important outcome of Appendix A is to illustrate that the average scattered field for infinite disordered metasurfaces with translation-invariant statistics comprises planewaves with an in-plane (parallel) wavevector identical to the incident planewave, $\mathbf{k}_{i,\parallel}$, in accordance with Snell's law. This characteristic holds universally across all space, encompassing the substrate, superstrate, and any thin layers deposited on the substrate. For a system illuminated by an incident planewave incident from the superstrate with a polarization $\hat{\mathbf{e}}_i$, the average electric field is given by

$$\langle \mathbf{E}_r(\mathbf{r}, +)\rangle_\infty = E_i\left[r_{sub}\hat{\mathbf{e}}_r + \rho\, \mathbf{J}(\mathbf{k}_{i,\parallel}, \mathbf{k}_{i,\parallel}, +)\, \hat{\mathbf{e}}_i\right] \exp[i\mathbf{k}_r \cdot \mathbf{r}], \tag{15a}$$

in the superstrate, and

$$\langle \mathbf{E}_t(\mathbf{r})\rangle_\infty = E_i\left[t_{sub}\hat{\mathbf{e}}_t + \rho\, \mathbf{J}(\mathbf{k}_{i,\parallel}, \mathbf{k}_{i,\parallel}, -)\, \hat{\mathbf{e}}_i\right] \exp[i\mathbf{k}_t \cdot \mathbf{r}]. \tag{15b}$$

in the substrate. Here the subscript "∞" denotes statistical averaging over an infinite metasurface. In Equations 15a and 15b, $\hat{\mathbf{e}}_r$ and $\hat{\mathbf{e}}_t$ represent the polarizations of the reflected and transmitted planewaves in the absence of particles, while $\mathbf{k}_r$ and $\mathbf{k}_t$ denotes their wavevectors, with $\mathbf{k}_{r,\parallel} = \mathbf{k}_{t,\parallel} = \mathbf{k}_{i,\parallel}$. Similarly, $r_{\text{sub}}$ and $t_{\text{sub}}$ represent the Fresnel coefficients of the layered substrate without particles. Note that the transmitted planewave is propagative or evanescent in case of total internal reflection.

With this information about the averaged field, one readily computes the coherent reflection and transmission coefficients for the monolayer (see the derivation of Equations A9 and A10 in Appendix A)

$$r_{ISA} = r_{\text{sub}} + \rho\big(\hat{\mathbf{e}}_r \cdot \mathbf{J}(\mathbf{k}_{i,\parallel}, \mathbf{k}_{i,\parallel}, +)\, \hat{\mathbf{e}}_i\big), \tag{16a}$$

$$t_{ISA} = t_{\text{sub}} + \rho\big(\hat{\mathbf{e}}_t \cdot \mathbf{J}(\mathbf{k}_{i,\parallel}, \mathbf{k}_{i,\parallel}, -)\, \hat{\mathbf{e}}_i\big). \tag{16b}$$

Due to interference effects, these coefficients may either exceed or fall below the values of $r_{\text{sub}}$ and $t_{\text{sub}}$ for the layered substrate without particles. These coefficients, which are pivotal in understanding the coherent scattering of waves from infinite metasurfaces exhibiting statistical translational invariance, are analogous to the celebrated Fresnel coefficients of planar interfaces.

It is worth remembering that the Jones matrix involves four complex-valued coefficients to address TE-TM cross-polarization conversion between the incident and refracted planewaves, with respective polarizations $\hat{\mathbf{e}}_i$ and $\hat{\mathbf{e}}_t$. For metasurfaces comprising rotationally invariant particles like nanospheres or nanodiscs, no cross-polarization conversion occurs, and only two complex coefficients are needed to characterize coherent scattering.

Equations 16a and 16b are derived by neglecting the interparticle interactions. They are accurate only for dilute metasurfaces (typically $\rho < 1~\mu\text{m}^{-2}$) and for small incident and scattered angles, $\theta_i$ and $\theta_s$ [96].

### Diffuse component of the scattered light

The diffuse contribution to the BSDF essentially represents the angle-resolved far-field statistically average speckle intensity $\langle |\delta \boldsymbol{E}_s|^2 \rangle$ scattered by the metasurface, see Equation 13. Again, owing to the ISA (or mean-field approach), the separation illustrated schematically in Figure 6 greatly simplifies the analysis. Following Appendix B, where we calculate the angle-resolved far-field scattered intensity $\mathcal{L}(\mathbf{k}_{s,\parallel}, \mathbf{k}_{i,\parallel})$ (refer to Equation B6 in Appendix B), we infer that diffuse contribution to the BSDF (measured in $\text{sr}^{-1}$) is

$$f_{s,\text{dif}} = \rho\; F\big(\mathbf{k}_{s,\parallel}, \pm, \mathbf{k}_{i,\parallel}, \hat{\mathbf{e}}_i\big)\; S_{r,\infty}\big(\mathbf{k}_{s,\parallel} - \mathbf{k}_{i,\parallel}\big) \frac{1}{\cos\theta_i \cos\theta_s}, \tag{17}$$

within the ISA. As expected, this contribution is directly proportional to the structure factor $S_{r,\infty}(\boldsymbol{q})$ of the infinite metasurface, and the form factor (in $\text{m}^2 \cdot \text{sr}^{-1}$) of the fictitious particle, see Figure 6b and Equation (9). The cosines account for the reduction in effective (projected) area due to the amount of light energy received and scattered by the surface at oblique angles.

It is worth noting that, similarly to the specular light, the ISA unrealistically predicts that the diffuse light scales proportionally to the density. Additionally, it is important to note that the BSDF diverges when the polar angles of the incident ($\theta_i$) and scattered ($\theta_s$) planewaves approach $\pm \pi/2$, (**C** is finite for these angles), whereas, in reality, disordered surfaces do not diffuse light and act as perfect mirrors at grazing incidences.

Despite these weaknesses, Equation 17 has a remarkable and intuitive consequence: it disentangles the distinct roles played by individual particles and structural correlations, thereby offering significant insights into the main two approaches for harnessing diffuse light. The first approach involves manipulating the particle itself and its interaction with the substrate to influence the form factor. The second approach entails controlling the constructive and destructive interferences that survives configurational averaging with correlated disorders. This control offers various structure factors through a vast landscape of arrangements [18], ranging from fully periodic structures to fully random uncorrelated disorders, passing through quasicrystals, weakly defective crystals, and hyperuniformity, as detailed in Appendix D.

*5.4 Advanced BSDF models of infinite metasurfaces*

The predictions of the ISA equations for the specular (Equation 16a and 16b) and diffuse (Equation 17) components of the scattered light do not consider interparticle interactions, leading to inaccuracies for significant incident angles (where $r_{ISA}$ and $t_{ISA}$ tend towards infinity as $\theta_i$ approaches $\pi/2$) and high particle densities — two situations where we expect interactions to be influential.

This Subsection presents a selection of recent models designed to incorporate particle interactions, primarily through mean field approaches. These models offer rapid computational speeds compared to full-wave simulations and demonstrate reliability even with relatively dense monolayers. However, they often necessitate validation through comparison with experimental data [97].

Specular component of the scattered light

There is a substantial body of literature on attributing effective constitutive permittivities and permeabilities to colloidal soft materials mean-field and effective medium approximations, such as Maxwell-Garnett or Bruggeman theories [38]. Despite their widespread use, these homogenization approaches have inherent limitations [98-101], particularly when applied to 2D monolayers. Their predictive force is considerably reduced for 2D monolayers, which are intrinsically anisotropic and often influenced by substrate proximity [81-83,85].

To achieve more precise predictions for 2D monolayers, mean-field approaches tailored to metasurfaces are necessary [38,88,102,103]. One such approach, known as the effective field approximation (EFA), considers the cumulative effects of multiple scattering on a statistical average basis. In the EFA, each discrete particle interacts with both the incident field and the averaged field scattered by all other particles. For monolayers, the statistically averaged field $\langle \mathbf{E}_s \rangle$ can be computed similarly to the ISA but without additional approximations, resulting in more accurate scattering coefficients than those derived from Equations 16a and 16b. For monolayers *in free space*, it can be demonstrated, either heuristically [97,104] or formally [96], that

$$r_{EFA} = \frac{\rho(\hat{\mathbf{e}}_r \cdot \mathbf{J}(\mathbf{k}_{i,\parallel},\mathbf{k}_{i,\parallel},+)\,\hat{\mathbf{e}}_i)}{1-\rho(\hat{\mathbf{e}}_t \cdot \mathbf{J}(\mathbf{k}_{i,\parallel},\mathbf{k}_{i,\parallel},-)\,\hat{\mathbf{e}}_i)}, \tag{18a}$$

$$t_{EFA} = \frac{1}{1-\rho(\hat{\mathbf{e}}_t \cdot \mathbf{J}(\mathbf{k}_{i,\parallel},\mathbf{k}_{i,\parallel},-)\,\hat{\mathbf{e}}_i)}. \tag{18b}$$

It is important to note that these formulas are applicable only in uniform backgrounds. However, they can be extended to layered substrates [96] using classical recursive relations for wave propagation in thin films [105]. The effective field approximation, being a higher-order approximation compared to the ISA, ensures that the coefficients exhibit correct behavior at grazing angles: $r_{EFA} \to -1$ for $\theta_i \to \pi/2$. This results in physically sound and quantitatively accurate predictions for moderately dense systems (typically, $\rho < 3$ µm$^{-2}$) even at relatively large incident angles (typically, $\theta_i < 60°$) [96].

Even greater accuracies can be achieved with models based on even higher-order approximations, such as the quasi-crystalline approximation. Within this formalism [106], the monolayer structure is modeled by the radial distribution function, which defines the likelihood of finding a particle at a specific distance from another particle. Several groups have utilized this approach to investigate the specular transmittance and reflectance of short-range ordered particulate monolayers [97,107-110] using state-of-the-art formulations. Further derivation details will not be provided here; interested readers may consult [110] and the references therein for comprehensive explanations. It is worth noting that a MATLAB program for computing specular reflection and transmission within the ISA, EFA and QCA approximations for monolayer in uniform backgrounds is provided in Appendix C.

Diffuse component of the scattered light

Diffuse light analysis can also benefit from refinements using the effective field or quasi-crystalline approximations to enhance the accuracy of the ISA model (Equation 17), particularly at high densities and grazing incident and scattered angles. While studies dating back to the 1980s have explored these approaches for 3D composite media [37,111,112], research on metasurfaces is scarce, especially compared to specular light investigations. Existing studies seem confined to monolayers of monodisperse nanospheres in homogeneous environments [113]. Nevertheless, for correlated disorders, these methods yield precise results for diffuse light across a wide range of densities, up to the maximum density of hexagonal lattices. This precision is attributed to the growing regularity observed at elevated densities, causing the metasurface to mimic the characteristics of a crystal, where the quasi-crystalline approximation is notably effective.

Alternatively, the predictive capability of the ISA model can be significantly improved by introducing a heuristic multiplicative correction term into the numerator of the final component of Equation 17 [79]. This term, rooted in physics, balances the cosines at grazing viewing and incident angles while ensuring reciprocity. Notably, it markedly enhances model precision, especially for moderate fill fractions, and appears capable of accurate predictions even for surface coverages nearing 10% [79]. Further refinement is expected with a renormalization of the form factor through mean-field approaches.

Despite their transparency in physics and utility in inverse design, the previous models share a common drawback: the form factor or multiple scattering corrections necessitate repetitive calculations across various frequencies, incidences, and polarizations, leading to inefficiency. Moreover, the resulting extensive 'lookup tables' obscure the underlying physics.

Recent advancements in enhancing the predictive capability of models for resonant particles leverage quasinormal mode theory [114] by approximating the form factor in Equation 17 with a sum of angle-resolved far-field scattered amplitudes, $\left(\frac{d\tilde{a}_s}{d\Omega}\right)_m$, computed for a few dominant resonance modes at complex frequencies [85]

$$F(\mathbf{k}_{s,\parallel}, \pm, \mathbf{k}_{i,\parallel}, \hat{\mathbf{e}}_i) \approx \left| \sum_{m=1}^{M} \alpha_m(\mathbf{k}_{i,\parallel}, \hat{\mathbf{e}}_i, \omega) \left(\frac{d\tilde{a}_s}{d\Omega}\right)_m (\mathbf{k}_{s,\parallel}, \pm, \hat{\mathbf{e}}_s) \right|^2. \quad (19)$$

This expansion separates the effects of the incident and scattered planewaves: the complex-valued expansion coefficients $\alpha_m$'s [115] have closed-form expressions and depend solely on the incident planewave, while the modal amplitudes $\left(\frac{d\tilde{a}_s}{d\Omega}\right)_m$ depend exclusively on the scattered planewaves. Replacing the highly multidimensional function, the form factor, with a weighted sum of a few far-field scattered modal amplitudes reduces computational loads and highlights physical quantities crucial for design purposes.

## 6. Bottom-up fabrication approaches

Numerous methods are available for creating disordered metasurfaces, and they can be broadly categorized into two main approaches: top-down and bottom-up.

Top-down patterning methods utilize advanced technologies such as electron-beam or optical lithography to precisely design the desired pattern. These technologies allow for the accurate placement of metaatoms at the nanometer scale, providing meticulous control over their size and shape [116]. For example, topology optimization [52] can enhance lattice coupling effects, and controlled rotational disorder (Figure 7c) can introduce optical activity without linear birefringence [117]. This level of mastery unlocks experimental opportunities not accessible to bottom-up approaches, offering the exploration of optical responses achievable through perfectly controlled disorder correlation. On the other hand, such approaches are limited by cost issues yielding metasurfaces with relatively modest lateral dimensions (typically up to 1 cm²). With the recent emergence of printing technologies based on soft lithography [118], such as nanoimprint lithography [119] or roll-to-roll nanoimprint techniques [120], the limitations concerning the necessary resolution and repeatability on a larger scale could be lifted.

In contrast, bottom-up approaches intrinsically prioritize scalability and cost-effectiveness, often allowing for nanofabrication without the need for clean-room facilities. While they may not achieve the same precision as top-down techniques in arranging nano-objects, their value lies in their ability to facilitate highly parallel nanofabrication on macroscopic surfaces (spanning several hundreds of square centimeters) with excellent conformity [116,121].

Therefore, exploring bottom-up approaches seems more beneficial, and we will concentrate on them in the following. Below, we outline various cost-effective fabrication methods, which are summarized in Table 1. It is important to emphasize that our goal is not to present an exhaustive list but rather to offer a foundation for further exploration. For more in-depth insights, recent comprehensive reviews offer substantial knowledge on the subject [11,122-125].

**Interrupted growth metal films.** Among the metasurfaces created using scalable and cost-efficient methods, one notable option is semicontinuous colloidal films of metal nano-islands. These films naturally emerge from deposition techniques interrupted during the initial growth of metal films before reaching the percolation threshold [29]. Initially explored for their huge potential in enhancing Raman scattering [30,31], these highly cost-effective films can be produced in large quantities by the glass industry, covering expansive areas up to tens of square meters.

Starting in the late 20th century, these semi-continuous films were employed to generate resonant optical surfaces and coatings, marking a significant historical trajectory in developing disordered surfaces with engineered plasmonic characteristics. This journey has given rise to surfaces that exhibit remarkable control over the reflection and transmission of light (Subsection 7.5) and have also demonstrated practical efficacy in fields such as optical biological and chemical sensing [126,127]. Additionally, the potential has been highlighted of harnessing diverse metal combinations to achieve even greater optical control over responsiveness [128]. Because the dimensions of these islands and the spaces between them are much smaller than visible wavelengths, semi-percolated metasurfaces function as coherent layers, absorbing incident light rather than scattering it. There are also alternative approaches, such as utilizing block copolymer templates for self-assembly [129,130], which enable the creation of monolayers with such tiny particles.

The process entails assembling pre-synthesized particles onto surfaces to create arrangements with larger particles that scatter light instead of absorbing it. Numerous techniques fall under this category, but we will focus on three general methods, emphasizing how "inherent disorder" can be harnessed to attain extensive and uniform metasurfaces.

**Bottom-up colloidal lithography**. This approach relies on the existence of colloidal solutions containing polystyrene, polymethyl methacrylate (PMMA), or silica nanospheres of nearly uniform size [131-134]. Once drop-casted and dried, these colloids form a monolayer. The latter then acts as a mask to imprint disordered patterns onto various underlying substrates. For example, it can be used to deposit metals onto surfaces or etch thin metallic films, resulting in the creation of disordered metasurfaces composed of nanoparticles [131,132], as well as surfaces featuring inverted structures like nanoholes in high-index films [24,24,135]. One noteworthy example in this category is hole-mask colloidal lithography, in which a thin metallic mask (made of materials such as Au, Cu, or Cr) with a pattern of (nano)holes is supported by a sacrificial PMMA layer [136]. Depositing various materials through such masks with a subsequent mask removal by a standard PMMA lift-off decorates the substrates with disordered metasurfaces. Several techniques contribute to this category [123,137]. They have found extensive use in producing optical metasurfaces characterized by correlated disorder for various applications, including optical chirality, magnetophotonics, thermoplasmonics, hybrid metal-molecular systems, and sensing in biological and chemical contexts [138-141]. Further, such metasurfaces can be directly transferred onto arbitrary substrates, as depicted in Figure 7a, or indirectly with 10 nm-thin (amorphous carbon) film carrier [121].

When using uncharged nanospheres, closely packed crystal arrangements are formed. These arrangements display well-ordered microscopic regions but lack long-range order on a larger scale due to crystal dislocations and defects.

**Table 1. Overview of Self-Assembly Techniques**

| Key Technique | Overview |
|---|---|
| **Drop Casting** | **Deposition on a substrate from a drying droplet with colloidal particles**. Self-assembly based on edgeward radial flow driven by evaporation, Marangoni recirculation and inter-particle & particle-substrate interactions.<br><br>**pros and cons**: Simple to implement but hard to control pattern formation; can form multiple rings at the edges and separate particle patches inside the drop due to depletion of particles in the middle. |
| **Evaporation in a Cell** | **Dispensation and evaporation of a particle suspension drop within a restricted area**, like a toroidal cell. Self-assembly driven by lateral capillary forces and convective influx.<br><br>**pros and cons**: Humidity and temperature control and wetting of the cell walls are required for better results. Defects formed due to particles sticking to the surface of the substrate before reaching the ordered region. |
| **Dip-Coating and Vertical Deposition** | **Deposition on a vertically held substrate in colloidal suspension**, where it is either withdrawn or kept stationary as the liquid evaporates or drains. Linear, continuous particle monolayer growth via convective transfer of the particles from the bulk of the suspension to the thin wetting film and interparticle and substrate-particle interactions.<br><br>**pros and cons**: Requires careful calibration of process parameters (e.g., temperature, humidity, withdrawal rate) to minimize defects. |
| **Drag-Coating** | **Continuous deposition of a liquid suspension** onto a horizontally translating substrate. Self-assembly mechanisms analogous to dip-coating, more compact implementation; easier control over parameters like temperature, but sensitive to substrate velocity and blade position.<br><br>**pros and cons**: Presence of regions of particle accumulation, nucleated from seeds, inhomogeneities and defects that need to be repaired via real time monitoring of the assembly process for order over large areas. |
| **Spin-Coating** | **Deposition of a colloidal suspension drop on a spinning substrate**, which spreads the liquid due to centrifugal force. Assembly dominated by shear-induced ordering for non-volatile solvents and inter-particle and particle-substrate interactions and capillary forces for solvents.<br><br>**pros and cons**: Need for careful adjustment of spinning protocols to prevent major defects; sensitive to parameters like particle concentration, solvent properties, and spinning speed. |
| **Electrostatic, Electrokinetic, and Electro-hydrodynamic** | **Adsorption of charged particles** on oppositely charged substrates or with an applied DC/AC voltage on electrodes in contact with the colloidal suspension. Driven by electrostatic attraction or electric field induced effects, like electrophoresis, dielectrophoretic, etc.<br><br>**pros and cons**: Electrostatic methods yield poor order and surface coverage; electrokinetic methods offer better control with proper protocol adjustment. |

Conversely, when employing charged colloids on charged substrates, one can leverage the interplay between inter-particle repulsion and the attraction of particles to the substrate. This interplay results in a short-range-ordered (amorphous) pattern characterized by correlated disorder [122]. The surface charges on the particles precisely control the spacing between particles. Consequently, the resulting pattern exhibits interparticle distances ranging from a few nanometers to several micrometers, determined by the dimensions of the dielectric colloids. As the sizing of beads like polystyrene is well-established, the transferred patterns, often arrays of nanodiscs or nanoholes in various materials, can be manufactured on a large scale with high precision and minimal aggregation.

Recently, this technique has found applications in a variety of contexts. For instance, it has been used to create energy-efficient plasmonic electronic paper that can achieve video-rate performance while displaying a broad range of vivid colors [143]. Additionally, it has been applied in the development of dynamically tunable conductive polymeric nanoantennas [144] and nearly hyperuniform optical metasurfaces [146].

**Self-assembly Nanoparticle Deposition.** Using this method, pre-synthesized colloidal nanoparticles, initially stabilized in a solution, are arranged on surfaces through techniques like blade-coating, drop-casting and dip-coating [143] (as shown in Figure 7b). A diverse range of nanoparticle shapes, including spheres, cubes, rods, rhomboids, stars, and others, are readily available commercially and can be organized into various arrangements, ranging from densely packed configurations to relatively dispersed patterns. The degree of control over these arrangements depends on the surface functionalization of individual particles. This functionalization is typically determined by the organic ligands that stabilize the nanoparticles during their wet synthesis. These versatile ligands serve multiple roles, acting as capping agents, responsive or conductive spacers, modifiers of wetting (acting as surfactants during assembly), or hydrophobic, electrostatic, or chemical linkers between particles.

Self-assembly deposition offers a straightforward method for creating large-scale conformal metasurfaces without requiring intricate nanopatterning or materials deposition steps. Various strategies fall into this category, including the drop-casting of nanoparticle solutions [143-148], assembly guided by (bio)chemistry principles [149], self-assembly utilizing block copolymer templates [150], gas-phase cluster beam deposition [84,151], or Langmuir–Blodgett assembly at air-water interfaces, accomplished by conducting the process in a water-filled trough with a movable barrier and pressure sensor [137,152]. It is worth noting that polymer-grafted nanoparticles can form a continuous layer with closely packed properties, even when the nanoparticles themselves do not exhibit a close-packed arrangement due to the presence of the polymer shell.

The assembly process can take two main forms. Firstly, it can occur directly on surfaces through techniques such as drop-casting from nanoparticle solutions. Alternatively, a pre-assembled layer of particles can be positioned atop a liquid-air interface, as demonstrated by the Langmuir-Blodget technique, or atop a sacrificial layer. The chemical linkages between particles can be intricate and extensive, often manifesting as substantial molecular assemblies. Block co-polymers, which serve as particle-particle spacers during quasi-disordered assembly on surfaces, are an example of this complexity.

**Block copolymers.** Block copolymers provide a well-established method for generating large arrays of densely packed features, typically ranging in size from 5 to 100 nanometers. These amphiphilic molecules usually consist of a hydrophilic block and a hydrophobic block, which form micelles when dissolved in polar solvents. On a substrate, they self-assemble into intricate patterns by segregating their opposing functionalities. This process leverages microphase separation, a phenomenon driven by the thermodynamic incompatibility between the blocks. The resulting morphologies—whether 1D, 2D, or 3D—are influenced by factors such as the degree of polymerization, the volume fraction of each block, and their interactions. Figure 7c illustrates a complex organization [153], with branching linear structures of nearly uniform width which form an interconnected network across the entire surface.

Block-copolymer nanopatterning serve as flexible templates for depositing metallic or semiconductor patterns [153] and is especially well-suited for crafting optical metasurfaces with controlled disorder on large and potentially non-planar surfaces [129,154]. It offers several advantages, including parallelism across the substrate, versatility in producing various features, and the ability to adapt to substrate size and curvature.

**Active metasurfaces.** Traditionally, disordered optical metasurfaces are fabricated from solid inorganic materials, whose geometry and light-scattering properties are difficult to modify once manufactured. The ability to dynamically tune their optical properties is constrained by the short interaction lengths in thin metasurfaces and the weak refractive-index variability in the visible spectral range. These limitations drive the search for alternative material platforms and tuning strategies for metasurfaces [155,156,150].

Active tuning of metasurfaces can be broadly classified into two categories: one that modifies the optical responses of the metaatoms near strong resonances, and another that relies on mechanical movement. The first approach typically involves altering the optical properties of the metaatoms or their surrounding environment, for example, through free-carrier injection, thermo-optic effects, electrochromism, phase-change materials, magneto-optics or structural transitions in various materials [17,157]. The second approach leverages macroscopic displacements, such as substrate stretching, or the actuation of micro-electromechanical systems (MEMS) to manipulate the metasurface properties [155-156, 150].

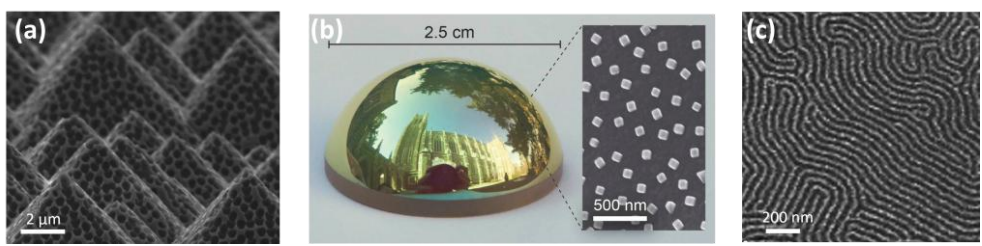

**Figure 7** Bottom-up nanofabrication approaches for disordered optical metasurfaces. **(a)** Anti-reflection coating of pyramid-textured Si surfaces with air holes obtained with colloidal lithography. **(b)** Bottom-up assembly of Ag nanocubes (right – top SEM view) on a curved macroscopic surface. **(c)** Tapping-mode AFM micrograph of block copolymers domains aligned over a length scale corresponding to several periods. Panel **(a)** used with permission of Royal Society of Chemistry, from "Highly conformal fabrication of nanopatterns on non-planar surfaces." Massiot *et al.*, Nanoscale, **8**, 11461-11466 [142], Copyright 2016; permission conveyed through Copyright Clearance Center, Inc. Panel **(b)** reprinted from Akselrod *et al.*, Adv. Mater. **27**, 8028-8034 (2015) [143]. Copyright Wiley-VCH Verlag GmbH & Co. KGaA. Reproduced with permission. Panel **(c)** reprinted from Progress in polymer science **32**, Darling, "Directing the self-assembly of block copolymers," 1152-1204 [153], Copyright 2007, with permission from Elsevier.

From the perspective of technology maturity, metasurfaces based on liquid crystals (LCs) or MEMS are among the most established candidates [158]. Both technologies make use of proven industry-standard and CMOS-compatible technologies to introduce active tuning capabilities. Resonant disordered metasurfaces with electrochromic polymers that can undergo reversible redox reactions (electron transfer) in the material may also become a valuable approach. The combination of resonance and disorder, combined with truly large scale, may be a valuable approach to expand the color palette of specular and diffuse hues, which are often limited to bland hues, in present smart windows.

However, the added complexity and associated increased cost-vs-benefit of active tuning present a challenge for the commercial deployment of active metasurfaces. Active disordered metasurfaces are less explored as their ordered counterparts, likely because applications willing to accept higher manufacturing costs, such as medical imaging, displays, and aerospace, are typically targeting the ordered metasurfaces. As a result, this review does not delve further into the development of active disordered metasurfaces, instead referring readers to other reviews for more in-depth discussions on the topic [155,159-160].

## 7. Applications

Optical interference is recognized as one of the foremost methods for enhancing the performance of contemporary optical devices. It plays a significant role across applications, leading to a significant overlap between the applications of periodic and disordered metasurfaces. This overlap often makes it intricate to ascertain which approach is superior with definiteness. Nevertheless, there are distinct features to consider when comparing periodic and disordered metasurfaces. In periodic arrangements, the collective behavior of all metaatoms within the structure generates highly enhanced optical interferences. On the other hand, in disordered arrangements, the coexistence of destructive and constructive interferences at the same spatial and spectral frequencies provides significant flexibility and richness, which is yet to be fully realized.

Early works on disordered optical metasurfaces can be traced back to Fraunhofer's work in 1887 on anti-reflective coatings using porous films [7]. Nowadays, disordered optical metasurfaces find a wide range of applications spanning various fields. These applications include Surface-Enhanced Raman Spectroscopy [10,31], achieving wide-angle absorption across a broad spectrum [84,143,152,161-163], reducing scattering cross-sections [164], implementing anti-counterfeiting measures [165], developing anti-reflection coatings [7,78,166] or sensors [167], manipulating wavefront [116,164,168], creating transparent displays [169,170] or new topological-design responses [171], and even producing optical chiral films devoid of linear birefringence [116].

These applications extend further into areas such as creating fade-resistant, angle-independent structural coloring [84,172-178], harnessing visual appearances [79,85,179-181], enabling rapid switching of plasmonic electronic paper [143], contributing to fields like quantum computing and optical communication, among others. Notably, disordered optical metasurfaces have a significant impact in solar radiation management, playing a crucial role in architectural and consumer products glazing for tasks such as interior cooling [9] and heating [180,182], solar-powered anti-fogging [183], heat protection using low-emissivity coatings [9,184-185], radiative cooling [186], optical encryption [187,188], light trapping in solar cells [135,189-191], and enhancing light extraction in light emitting diodes [45,192-196].

In the subsequent Subsections, we refrain from an exhaustive review of efforts undertaken in these diverse realms, given that a comprehensive review has been recently presented [16]. Instead, our focus shifts towards the selection of recent endeavors that illustrate the promising facets of applications. Through this, we aim to provide readers with an understanding of how to put into practice the theoretical constructs outlined in Sections 3 and 5. We try to emphasize the potential of design-induced resilience to minor imperfections in features, size, and spacing—an almost inevitable reality in large-scale manufacturing—, for instance for the creation of vibrant colors despite the polydispersity of resonant metaatoms in terms of size and shape, as facilitated by bottom-up fabrication techniques [85,176].

## 7.1 Tailored optical diffuser

A diffuser serves as a device that essentially strongly scrambles the incident wavefronts and reduces its spatial coherence. For example, when a highly spatially coherent laser beam encounters a diffuser, the light emerging from the diffuser may no longer have the characteristics of a beam, but rather propagate in a wide range of directions. Optical diffusers play a crucial role in numerous photonic and optoelectronic applications [197], including illumination [192], displays [169-170,198-199], imaging systems [200-203], photovoltaics [22,190], wavefront shaping [116,204], and advanced applications like generating optical physical unclonable functions for cryptography and optical random number generators [197].

The detailed characteristics of the scattered intensity can differ substantially among different types of diffusers. For instance, certain diffusers, like sandblasted optical surfaces, scatter partially within a limited angular cone, exhibiting a bell-shaped scattering distribution centered around the specular direction (Figure 8a).

A noteworthy and practically important category is that of Lambertian volume diffusers, introduced by Johann Heinrich Lambert [205]. These diffusers appear uniformly bright from all observation directions, adhering to Lambert's cosine law, with radiant flux per unit solid angle diminishing as the observation angle approaches 90° (Figure 8b). Lambertian diffusers are valuable for reducing glare and eliminating areas of intense brightness that can be discomforting to the human eye. They find applications in photography, lighting design, and display technology, where diffuse illumination is often advantageous.

In contrast to traditional diffusers developed in the past, diffusers based on metasurfaces introduce new possibilities in design and can facilitate the incorporation of novel functionalities [197]. The increased design flexibility offered by metasurface-based diffusers becomes apparent when examining the BSDF models outlined in Section 5.

Initially, the metaatom density $\rho$ serves as a critical parameter, allowing the adjustment of the balance between diffused and specular light. It is important to note that the model in Section 5 operates within the framework of the ISA, and the predicted linear increase in diffuse light with density holds true for small densities only. Generally, once the density surpasses a few metaatoms per square micron, the proportion of diffuse light saturates and generally decreases at higher densities.

Another degree of freedom involves engineering the form factor, where various aspects can be implemented. For example, achieving far-field directional scattering relative to the metaatom orientation (Figure 8c) rather than the metasurface normal direction is possible by controlling interference among several dipolar modes [206-209]. This interference process can also enhance the efficiency of forward diffusion. While Lambertian volume diffusers scatter light uniformly, it is conceivable to implement diffusers that predominantly scatter light forward by overlapping electric- and magnetic-dipole modes. This approach was successfully tested numerically in [210], where the authors observed that, in an average sense, nanoparticles retain their individual Kerker behavior in the presence of random interparticle coupling, despite the stringent requirement placed on both the amplitude and phase of the electric- and magnetic-dipole modes [211]. The form factor also allows for the use of resonant metaatoms, which selectively scatter light within a narrow spectral window and fade away outside this window [212]. Resonant approaches are advantageous for achieving perfect Lambertian scattering with a response independent of polarization and incident angle when spectral selectivity is required.

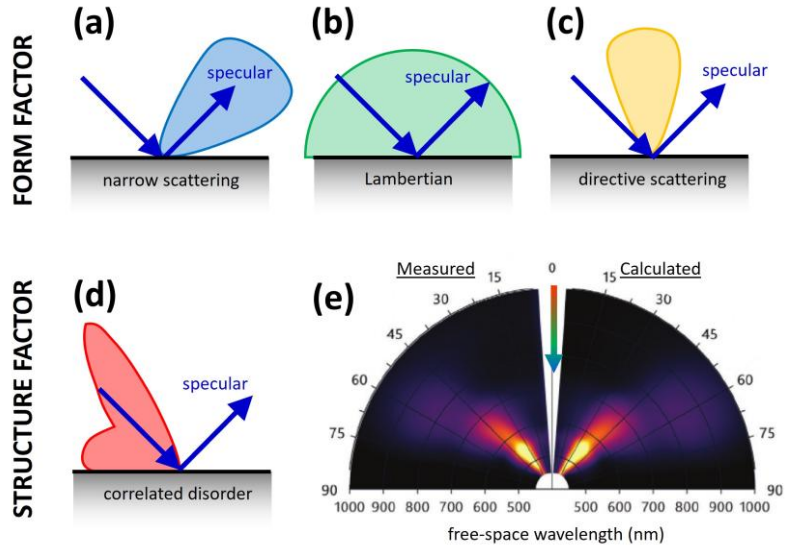

**Figure 8.** Metasurfaces enrich the arsenal of scattering regimes. **(a)** Bell-shaped, relatively narrow scattering distribution centered around the specular direction is obtained with classical diffusers. **(b)** Lambertian scattering is achieved with volume diffusers. **(c)** Directive scattering fixed by the metaatom orientation is achieved with metaatoms combining several multipolar contributions. **(d)** With correlated disorder, light is not scattered around the specular direction. **(e)** experimental verification with visible light and TiO$_2$ nanodiscs. Panel **(e)** reprinted from Piechulla *et al.*, Adv. Opt. Mater. **9**, 2100186 (2021) [23]. Copyright Wiley-VCH Verlag GmbH & Co. KGaA. Reproduced with permission.

The third influential parameter is the structure factor (Figures 8d-e). By introducing short-range order or hyperuniformity, it becomes possible to significantly suppress the diffuse light around the specular direction. Referring to Figure 6 and defining the quenching zone as $S < 1$, or equivalently $|qa| < 2\pi A$ with $A$ a dimensionless parameter slightly smaller than unity, it is deduced that light predominantly scatters outside an angular cone

around the specular direction, and the cone apex $\theta_q$ is given by $\sin(\theta_q) < A\lambda/a$. Considering that $a$ is the inverse of the square root of the density ($a = \rho^{-1/2}$), a grating-like equation emerges, indicating that the quenching zone is proportional to the average distance between nearest neighbors and wavelength. This unique scattering regime (Figure 8d) has been experimentally demonstrated with various high-index metaatoms [23,79] and across a broad range of incident angles. Figure 8e provides a compelling demonstration for normal incidence.

## 7.2 Light absorption in solar cells

One major challenge in global energy supply is improving the material efficiency of solar cells by reducing material usage while maintaining or enhancing conversion efficiency. Conventional absorber materials like crystalline silicon (c-Si) and promising newcomers like perovskites possess intrinsic optical characteristics that necessitate attention [213-219]. These materials exhibit notable back reflection issues at the top interface, primarily because of their high refractive indices, while absorption is low for energies close to the bandgap, particularly for indirect semiconductors like c-Si but also for thin layers of direct semiconductors.

Creating effective anti-reflection and light-trapping mechanisms for thin solar cells presents a significant challenge. This is because they need to exhibit a wide tolerance for both angular and spectral responses, a necessity driven by the Earth's rotation and the sun's broad spectrum, where approximately 90% of its radiant power falls within the wavelength range of 300 to 1500 nm.

For crystalline silicon, traditional thick-layer light management methods, like the successful pyramidal interface [220], are not suitable for ultra-thin (≤ 10 µm) solar cells. Other scalable fabrication approaches applicable to thinner absorber layers, achieved through processes such as metal-assisted wet chemical etching, fs-laser irradiation, or electrochemical etching and similar methods [221-225], effectively suppress reflection (antireflection) and enhance optical path length (light trapping), resulting in so-called black silicon with absorption levels close to the $4n^2$ limit, the reference standard deduced from the Lambertian scattering interface [205,226-229]. However, these fabrication methods usually introduce numerous surface defects, and the natural expansion of the textured surface area relative to a flat surface inevitably results in a higher density of defects per unit area. This frequently results in severe electronic degradation at the interface, including a significant recombination of photo-generated electron-hole pairs [223,225], which counteracts the advantage of the augmented photocurrent unless tailored electronic passivation techniques are implemented [224-225,230].

Certainly, solar cells pose a complex challenge in optoelectronics, underscoring the importance of considering factors beyond optics alone, as emphasized in a recent review by Massiot et al. which compiles the latest breakthroughs in ultra-thin c-Si, GaAs, and CIGSe solar cells [215]. Despite the implementation of advanced light-management techniques for which optical simulations predict significantly improved performance, the actual short-circuit currents achieved in experiments barely exceed the double-pass absorption limit, akin to having a basic rear side mirror in an unstructured cell. In fact, only 3 out of 17 experimental studies on ultra-thin c-Si solar cells were able to surpass this limit.

This discrepancy suggests that forthcoming strategies for light management in ultra-thin solar cells should aim to preserve the electronic properties of the absorber material to the greatest extent feasible. Ideally, the absorber material structural integrity should be conserved, keeping it in a flat configuration, while metasurfaces are employed on either the front, back, or even both sides [213,231-233]. A front-side metasurface would address anti-reflection concerns and potentially introduce light-trapping mechanisms, while a rear-side metasurface would focus exclusively on light trapping. Employing metasurfaces on both sides allows for the separation of anti-reflection and light-trapping functions, which could be advantageous since they serve distinct purposes with unique requirements.

Figure 9a depicts the overall criteria one should anticipate from a rear-side metasurface (depicted in violet) to prevent the easy escape of light from the absorbing layer (shown in gray) [190,234-235]. The metasurface function should include the suppression of scattering into angles within the emission air cone. Additionally, it should amplify scattering into angles greater than the angle of total internal reflection at the top interface of the absorber to maintain light within the absorber, stimulating its guided modes and thus effectively enhancing the optical path length. Lastly, it is crucial to avoid scattering into evanescent waves characterized by large in-plane wavevectors, as these waves do not propagate back into the absorber but rather remain confined to the rear side. This confinement could lead to potential high parasitic absorption in non-transparent rear side nanostructures. All these requirements are graphically depicted in reciprocal space in Figure 9b, where $k_x$ and $k_y$ represent the $x$- and $y$-components of the in-plane wavevector.

For a periodic metasurface, the grating constant plays pivotal role to achieve such reciprocal space engineering. Periodicity causes the structure factor $S_r(\boldsymbol{q})$ (Equation 14 in Subsection 5.2) to compact into a set of Dirac delta distributions, which channels scattering into the well-known diffraction orders. Due to the ease of experimentally manipulating the grating constant, a plethora of approaches utilizing periodic metasurfaces have been introduced [215,232,236-242]. Nevertheless, their light trapping performance usually falls short when compared to random interfaces, such as black silicon, and it turns out that reaching the 4n² limit proves challenging in practice.

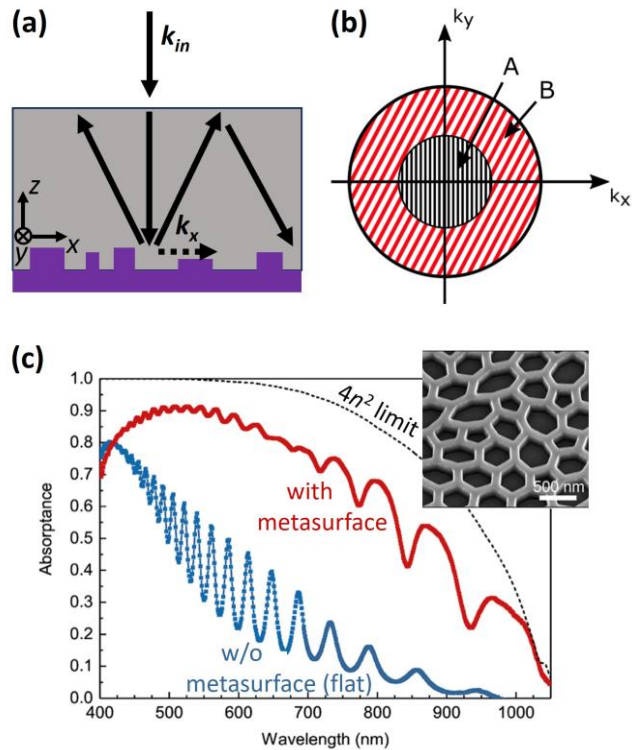

**Figure 9.** Hyperuniformity for light absorption. (**a**) A light-trapping metasurface (violet) at the rear side of a solar cell absorber (gray) ideally scatters weakly absorbed incoming light into directions for which total internal reflection occurs at the frontside. (**b**) Reciprocal momentum-space representation for the parallel wavevectors of evanescent and propagative planewaves scattered by the metasurface. Zone A: emission cone in air ($q = \left(k_x^2 + k_y^2\right)^{1/2} < \omega/c$), zone B: emission cone in the absorbing layer ($q < n_a\omega/c$) with $n_a$ the refractive index of the layer. The planewaves with a parallel momentum modulus greater than $n_a\omega/c$ are evanescent and do not propagate back into the absorber but are bound to the rear side with potentially high parasitic absorption, such as surface plasmon polaritons. (**c**) An ultra-thin 1µm-thick crystalline silicon membrane exhibits a remarkable 65% increase in absorption (red curve) across a broad spectrum from 400 nm to 1050 nm, thanks to the implementation of a hyperuniform metasurface (inset), in contrast to the flat surface lacking a metasurface (blue curve). Panel (**b**) reprinted with permission from [190] © Optical Society of America. Panel (**c**) adapted from [22], CC BY 4.0.

An alternative, yet powerful, strategy to achieve the targeted scattering distribution is the use of correlated disorder [161,189-191,243-245]. This strategy encompasses two complementary methodologies. In the first approach, the high index absorbing film with a random patterning is seen as a 2D platform that facilitates the formation of a myriad of leaky resonance modes [114]. These modes, once stimulated by broadband light, boost absorption similarly to grating waveguides. Correlated disorder allows to adjust the density and spectral distribution of spatial modes, as well as their coupling in k-space [161,189]. The second approach, illustrated in Figure 9, relies on the ISA framework. Starting from the constant structure factor $S$ of a completely random (Lambertian) interface and introducing spatial correlations, e.g., via an average distance between scattering elements, it is possible to tailor $S$ such that scattering within the emission air cone is suppressed but emission outside is enhanced. Correlated disorder is a highly appealing concept for light trapping, offering the strengths of both random and periodic structures, e.g., the unique combination of operation across a wide spectral and angular spectrum, while also providing the ability to selectively enhance or suppress scattering. Notably, the latter enables the resonant excitation of guided modes, markedly boosting absorption, like in the first approach.

Upon revisiting the review by Massiot et al. [215], it is clear that two of the three highest-performing ultra-thin crystalline silicon (c-Si) solar cells utilize periodic metasurface light-trapping strategies [216,217], while the third relies on a correlated disorder approach [208]. Notably, the solar cell employing correlated disorder achieves the greatest relative photocurrent enhancement compared to single-pass absorption, with a +79% increase, whereas the periodic metasurfaces show improvements of +34% and +43%, respectively. This substantial performance gain highlights the promising potential of correlated disorder metasurfaces for advanced light management in ultra-thin solar cell designs.

Clearly, the desired reciprocal-space illustration in Figure 9b bears similarities to the one shown for the structure factors of stealthy hyperuniform arrangements [246] discussed in Appendix D. Indeed, some of the most effective light trapping configurations featuring correlated disorder appear to exhibit (nearly) hyperuniform disorder, although they were not explicitly identified as such [161,189-191,244,247]. The exploration of hyperuniform disorder for light trapping is still in its early stage [22,233,248-249]. A remarkable result was the demonstration of over 65% light absorption (Figure 9c) in a 1µm-thick silicon slab achieved through a carefully engineered hyperuniform disordered metasurface (Figure 9c) [22]. This metasurface was tailored for efficient light coupling into the available slab modes, particularly under normal incidence and additional experiments are essential to assess its light-trapping capabilities under oblique incidence.

However, the practical application of these advanced designs in solar cells is limited, partly because of the sophisticated fabrication techniques required, such as electron beam lithography. Given the photovoltaic industry's emphasis on cost-effectiveness, only those hyperuniform disordered structures that can be economically produced using techniques such as colloidal lithography or (roll-to-roll) nanoimprint technologies are considered feasible [135,250-251]. To the best of our knowledge, only a single study has successfully applied a hyperuniform disordered metasurface to a functional large-scale industrial crystalline silicon solar cell so far [248]. By exploiting colloidal lithography as a scalable bottom-up fabrication method, it was demonstrated that this process has no detrimental effects on the electronic properties of the cell and leads to enhanced photocurrent (+5.1%) due to light trapping and suppressed front side reflection.

### 7.3 Light extraction

A similar discussion can be initiated regarding the dual challenge of extracting light emitted from sources embedded within intricate thin-film stacks. Consider, for instance, organic light-emitting diodes (OLEDs), which find widespread application in electronic device displays and have the potential to revolutionize lighting design by introducing concepts like lighting planes instead of traditional localized lamp sources. Achieving high efficiency is crucial for reducing energy consumption in lighting installations.

However, in conventional planar OLEDs, assuming a typical planar device structure as illustrated in Figure 10a, only approximately 20% of the power emitted by the organic molecules is emitted into free space [252-254]. This is primarily because various undesired modes within the system are excited due to the broad distribution of in-plane k-vectors resulting from small dipolar emitters like molecules. Consequently, power is lost by coupling light into lossy waveguide modes, decaying into surface plasmon modes at the metallic electrode, or trapping light via total internal reflection into the modes of the glass substrate.

In contrast to the situation in solar cells, where the objective is to redirect as much light as possible out of the emission cone, here, the goal is to direct as much light as possible into the emission cone, specifically within angles smaller than the angle of total internal reflection.

Numerous methods employing nanostructures to enhance light extraction have been proposed. These approaches can generally be classified into two categories: external [45,255-258] and internal [252,259-264] scattering layers. External scattering layers are applied at the substrate/air interface. They are independent of the actual light-emitting structure, and are thus more easily integrated into existing devices [45,258-259]. However, they primarily impact the loss channel related to light trapping in substrate modes. On the other hand, internal scattering layers are integrated in close proximity to the active layers, where waveguide or surface plasmon modes are excited, effectively addressing these loss channels by enabling their conversion into substrate modes and free-space channels.

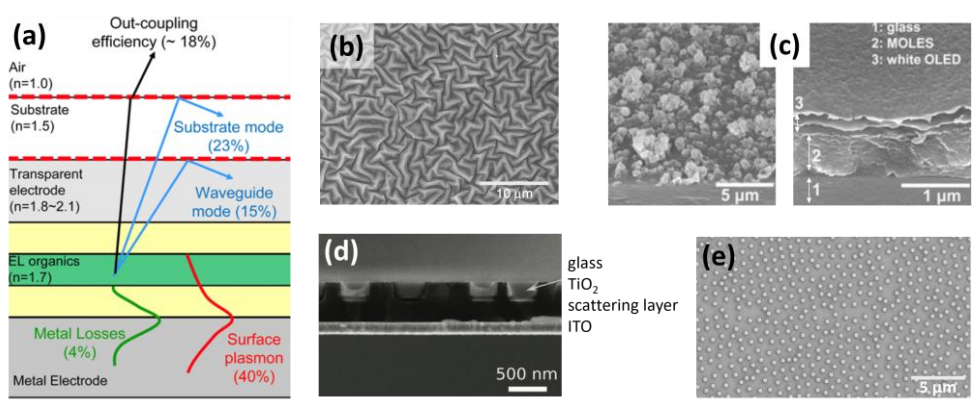

**Figure 10.** Diverse disordered metasurface configurations for enhanced light extraction. **(a)** Illustration of a typical planar multilayer OLED, featuring an electroluminescent organic layer (green) situated between two injection and transport layers (yellow). These layers are connected to metal and transparent electrodes and are applied onto a much thicker glass substrate. **(b)** Representation of a corrugated polymer substrate designed for white light OLEDs. **(c)** A straightforward yet highly effective disordered metal oxide-based internal scattering layer, depicted before (left) and after (right) planarization and deposition of a white OLED multilayer stack. **(d)** Cross-sectional view of an internal scattering layer composed of a nearly hyperuniform disordered monolayer of $TiO_2$ nanodiscs. **(e)** An external scattering layer tailored for LEDs, featuring an almost hyperuniform disordered monolayer of $TiO_2$ nanodiscs. Panel **(a)** reprinted by permission from Springer Nature: Hong *et al.*, Electron. Mater. Lett. **7**, 77-91 [254], Copyright 2011. Panel **(b)** reprinted from [262], CC BY 4.0. Panel **(c)** reprinted from Kim *et al.*, Adv. Funct. Mater. **24**, 2553-59 (2014) [261]. Copyright Wiley-VCH Verlag GmbH & Co. KGaA. Reproduced with permission. Panel **(d)** reprinted from Piechulla *et al.*, Adv. Opt. Mater. **11**, 2202557 (2023) [264]. Copyright Wiley-VCH Verlag GmbH & Co. KGaA. Reproduced with permission. Panel **(e)** adapted from [257], CC BY 4.0 (https://creativecommons.org/licenses/by/4.0/).

The conversion of the waveguide or surface-plasmon modes into substrate modes using an internal layer, or the conversion of trapped substrate modes into free-space modes with an external layer, can be effectively achieved through the use of periodic scattering layers, as documented in previous works [256,259-260]. However, employing periodicity introduces pronounced angular and spectral dependencies, resulting in undesirable effects such as butterfly-like emission patterns and chromatic aberrations. These detrimental effects can be circumvented by employing disordered structures

[45,193,252,257,261,262,264]. The primary objective is to carefully engineer disorder so that metasurfaces can effectively provide the desired k-space momentum towards a uniform enhancement of mode conversion within the emission cone. A growing area of interest revolves around disordered metasurfaces that exhibit strong short-range correlations [257,264-266]. These metasurfaces are particularly intriguing as they offer innovative opportunities for manipulating k-space for enhanced light extraction (Figure 10b-e), especially when combined with highly scattering elements such as high-refractive-index or metallic particles, instead of relying on corrugated interfaces.

### 7.4 Color and appearance coatings

For a long time, researchers have studied metasurfaces that utilize photonic modes to create grating-like effects, which are observable only at certain illumination angles, within specific spectral ranges or polarizations. The manufacturing of such metasurfaces has reached an advanced stage of industrial development, with optical security tags being a prime example.

Over the last decade, there has been a growing interest in angle- and polarization-independent colors produced by metasurfaces with Mie or plasmonic resonances. These metasurfaces offer the potential for creating color coatings that are not only lighter and more resistant to fading but also could be less harmful to the environment compared to traditional chemical dyes and colorants. They exhibit remarkable color vividness and saturation [267], alongside outstanding stability against chemicals, extreme environmental conditions, and prolonged exposure to bright light. Furthermore, these metasurfaces can achieve resolutions beyond the limits of diffraction, which is especially beneficial for display technologies [268]. For a deeper understanding, refer to the detailed reviews in [269-271].

However, these impressive achievements mostly rely on periodic structures that are sensitive to viewing angles and require costly, low-throughput nanofabrication methods. It is crucial to remember that even slight discrepancies in the shape or size of the metaatoms can hinder the achievement of pure colors. For instance, a slight variation of just 10 nm in size can lead to a resonance frequency shift by 10-20 nm. Consequently, the combined goals of low manufacturing costs, vibrant colors, and angle insensitivity represents a significant challenge. As a result, there are currently no structural paints available on the market.

Semicontinuous films of nano-islands are cost effective; however, alone, they cannot provide vibrant colors. They combine localized plasmonic resonances at individual islands, hot spots in the gaps between islands, and delocalized modes. This multiscale confinement and the size polydispersity results in a very broad plasmonic resonance, which unfortunately narrows the potential applications of nano-island films for color production. So far, the only application for these films has been to alter the reflectance of conventional Ag mirrors — which are known for their very high and wavelength-independent reflectance — by introducing a warmer, peach-toned silver effect for a more pleasing aesthetic when we look ourselves in the mirror [272].

A novel approach has recently been developed [176]. It slightly differs by introducing an additional, extremely thin alumina layer, roughly 10 nm in thickness, between the metal surface and the nano-islands. This adjustment adds a new way to manipulate light absorption, thanks to gap plasmons that are confined within the alumina layer (Figure 11a1). By varying the size of the nano-islands, a broad spectrum of colors can be achieved, even in the presence of significant size variations among the nano-islands (leftmost inset of Figure 11a2).

Promisingly, the nano-island films, which can be evaporated over large areas, can also be flaked off and mixed in commercial paint oils. This process paves the way for the development of ultra-light plasmonic structural color paints (Figure 11a3) which can be applied to variously textured surfaces (rightmost inset in Figure 11a2). It is worth acknowledging that the inherent diversity and randomness of this method do introduce some broadening of the optical resonances, which slightly diminishes the brilliance and saturation when compared to periodic metasurfaces. Despite this, the potential for low-cost production opens up exciting avenues for future developments in this domain.

While the presence of gap plasmons within the alumina layer offers a plausible explanation for the resilience of the color to size polydispersity, it is crucial to consider other contributing mechanisms. Particularly, the generation of subtractive colors through the absorption by the monolayer, perceived as a uniform film, marks a departure from the scattering-based coloration seen with larger nanoparticles in periodic metasurfaces.

Manipulating the color of objects through coatings that utilize Mie or plasmonic resonances represents just the initial phase in achieving a control over our perception of these objects. The visual attributes of objects extend beyond color and encompass factors such as gloss, haze, transparency, texture, illuminant spectrum, and much more [273].

In Figure 11b, we illustrate the profound impact of nanoscale features on our perception of objects. Specifically, we examine the influence of surface roughness on the perceived colors of metal films perforated by nanohole arrays. The photorealistic images are generated through FDTD computations of the zeroth order specular reflectance of the array, followed by rendering [274]. They underscore that nanoscale surface roughness can influence the perceived colors of patterned thin films in ways that are difficult to predict.

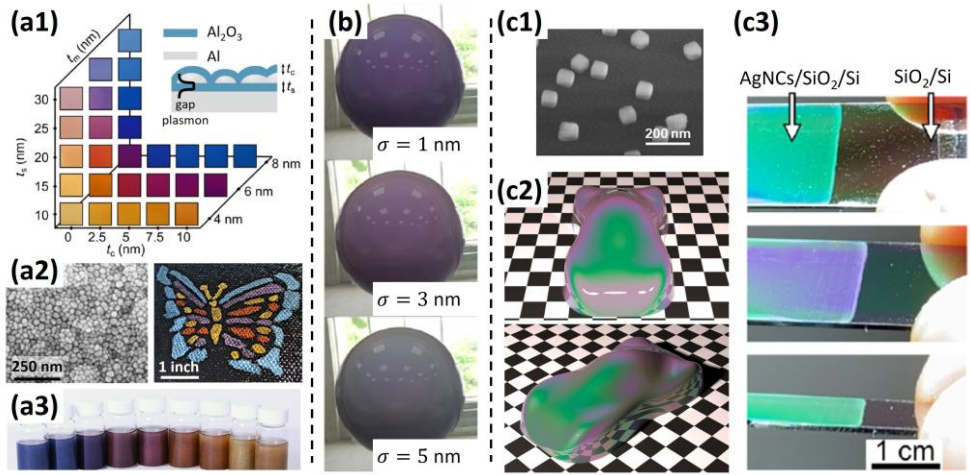

**Figure 11.** **(a1)-(a3)** Low-cost color coatings with nanogap plasmons formed by Al nano-islands on $Al_2O_3$/Al substrate. The metasurfaces offer quite vivid hues **(a1)** by tuning the effective film thickness $t_m$ and the spacer and capping layer thicknesses, $t_s$ and $t_c$. The leftmost inset in **(a2)** shows a SEM picture of a metasurface with an effective thickness of $t_m = 5$ nm and the rightmost inset is a photography of a black canvas painted with oil–based plasmonic paints **(a3)** obtained by dispersing metasurface flakes in commercial paint oils. **(b)** Impact of metal roughness on perceived color is illustrated using rendered images of a macroscopic sphere coated with a 20-nm-thick Ag film featuring periodic hole arrays and increasing metal roughness. Diffuse light is not considered in the rendering, which explains why the glossiness of the surface remains constant despite an increase in the root mean square ($\sigma$) of the surface height amplitude. **(c1)-(c3)** Two-color iridescence exhibited by a monolayer of Si nanocubes **(c1)** on a $SiO_2$/Si substrate. Rendered images **(c2)** and naked-eye observations **(c3)** consistently reveal only two distinct vivid colors (hue difference of 160°) across all viewing and lighting directions. Panel **(a1)-(a3)** from Cencillo-Abad *et al.*, Sci. Adv. **9**, eadf7207 (2023) [176]. Reprinted with permission from AAAS. Panel **(b)** reprinted with permission from [274] © Optica Publishing Group. Panel **(c1)** and **(c3)** adapted with permission from Agreda *et al.*, ACS Nano. **17**, 6362-6372 [85], Copyright 2023 American Chemical Society. Panel **(c2)** adapted with permission from Springer Nature: Vynck *et al.*, Nat. Mater. **21**, 1035-1041 [79], Copyright 2022.

The concept of appearance is intricate and has primarily been explored by designers and researchers in computer graphics within the realms of fine and applied arts. Yet, it has received limited attention from researchers in nanophotonics. This lack of exploration prompts the inquiry into the generation of novel appearances using nanostructures that diverge from natural configurations.

Various degrees of freedom can be harnessed, and only recently have initial responses been provided for metasurfaces composed of monolayers of resonant nanoparticles using the BSDF models presented in Subsection 5.4 [79]. The manipulation can involve either the form factor or the structure factor. Figure 11c exemplifies a novel appearance achieved through the former approach. It shows rendered images of a disordered monolayer of silver nanocubes deposited on a silicon substrate covered with a thin silica film (Figure 11c1). A unique iridescence emerges through the hybridization of dipolar and quadrupolar nanocube resonances with their mirror images in the silicon substrate. Unlike classical thin film iridescence, where colors change progressively with varying viewing angles, the diffuse light, in this case, exhibits only two remarkably distinct vivid colors — green and purple as shown in Figures 11c2 and c3 — across all illuminating and viewing angles [85].

Visual appearance can also be harnessed by metasurfaces implementing correlated disorder. As detailed in Appendix D and Subsection 7.2, introducing short-range order leads to a notable reduction in diffuse light, particularly in the vicinity of the specular direction. Essentially, this reduction results in coatings consistently appearing dark when viewed in the direction of the specular reflection. On curved surfaces, the specular direction undergoes non-uniform changes due to curvature. Dark areas thus appear on uniform coatings and the areas deform and glide across the surface when the observer move [79].

### 7.5 Transparent metasurface for augmented reality

A passive transparent display is a type of see-through screen or visual interface that does not emit its own light but relies on external light sources to display information. This capability to show graphics and text on a see-through screen while allowing the background scenery to be visible at the same time has a wide range of valuable applications, such as heads-up displays, smart windows, and information panels.

In Subsection 5.3, we explained that there are primarily two key methods for effectively controlling diffuse and specular light, whether it is through reflection or transmission. These methods involve altering either the form factor, which relates to the far-field radiation pattern of the metaatom, or the structure factor, which concerns the statistical arrangement of particles. Going forward, our goal is to illustrate the practical implementation of these two approaches by using the example of designing metasurfaces that serve as passive transparent displays. Additionally, this will provide an opportunity to introduce interesting characteristics of cascaded disordered metasurfaces made up of two or more monolayers.

One notable advantage provided by form factor engineering is the potential to achieve a wavelength-specific response through metaatoms that possess distinct and precise resonances. These metaatoms selectively scatter light of a particular wavelength while displaying minimal responsiveness to other wavelengths [169,198,212]. Consequently, when we apply these resonant nanoparticles onto a transparent substrate and project images using the resonant wavelength, the metasurface scatters the majority of the projected light while preserving high transparency to ambient light with a broad spectrum (Figure 12a). This concept was exemplified in [169], where silver nanospheres embedded in a polymer matrix were utilized to create a blue color display. This resulted in a 100-nanometer wide extinction peak, centered around the resonance wavelength of 460 nm, consistent with simulation results.

To develop a comprehensive full-color display, it becomes possible to use three distinct types of nanoparticles, each specialized in scattering light at one of the three desired colors: red, green, and blue. The primary challenge lies in achieving sharp resonances while simultaneously maintaining significant transparency for wavelengths outside of these specific resonant points.

Transparent displays can also be realized by leveraging the structure factor, utilizing two disordered monolayers. A notable characteristic of multi-layer metasurfaces is their distinctive behavior regarding specular light in transmission versus reflection. This contrast becomes clear when we examine the expressions governing the diffuse and specular contributions to the scattered light in Equations 12 and 13, and consider that **q** represents the difference in wavevector between the scattered and incident directions. In the case of monolayer metasurfaces, like the ones investigated thus far, all the metaatoms are situated in the same plane, say $z = z_0$, and the term $q_z z_m$ in the exponential expression $\exp[i\mathbf{q} \cdot \mathbf{r}_m]$ found in Equations 4 and 6 remains constant.

This property no longer holds true for multi-layer metasurfaces. In this scenario, $z_m$ becomes a random variable that can assume various distinct values, with two values being relevant for a two-layer metasurface discussed hereafter. To grasp how this property can be utilized in designing a transparent display, let us examine the expression for specular intensity, which is proportional to $N^2|\langle\exp[i\mathbf{q} \cdot \mathbf{r}]\rangle|^2$, as predicted by the ISA model, see Equation 12. It is apparent that in transmission ($\mathbf{k}_s \equiv \mathbf{k}_t = \mathbf{k}_i$), $\mathbf{q}$ equals zero, and the specular intensity is simply $N^2$, which is the square of the total number of metaatoms in the two monolayers. When dealing with non-resonant metaatoms, the specular light predominates, and the bilayer metasurface behaves like a transparent window.

In contrast, when considering reflection ($\mathbf{k}_s \equiv \mathbf{k}_r = -\mathbf{k}_i$), $\mathbf{q}$ becomes $-2\mathbf{k}_i$, causing the exponential term $\exp[i\mathbf{q} \cdot \mathbf{r}]$ to transform into $\exp[-2ik_{i,z}\hat{\mathbf{z}} \cdot \mathbf{r}]$, where $k_{i,z}$ represents the z-component of the incident wavevector. The statistically averaged value of this exponential term will strictly become zero if each layer contains the same density of metaatoms and if the separation distance $t$ between the layers results in a $\pi$ phase-shift difference, meaning $2k_{i,z}t = \pi$. Under this specific condition, the specular contribution to the scattered light disappears, and the diffuse light takes over, causing the bilayer metasurface to function as a diffusive screen.

In practical terms, the challenge lies in achieving a broadband $\pi$ phase-shift difference that is ideally minimally influenced by the angle at which incident light approaches. Recently, a successful broadband cancellation of specular reflection has been demonstrated across the entire visible spectrum through a precise combination of dispersion properties. This combination involves a thin film stack comprising two monolayers of metal metaatoms and two dielectric thin films with meticulously designed thicknesses [170]. The effectiveness of this asymmetric transparent display, which primarily scatters reflected light while maintaining the integrity of transmitted light, is illustrated in Figure 12b.

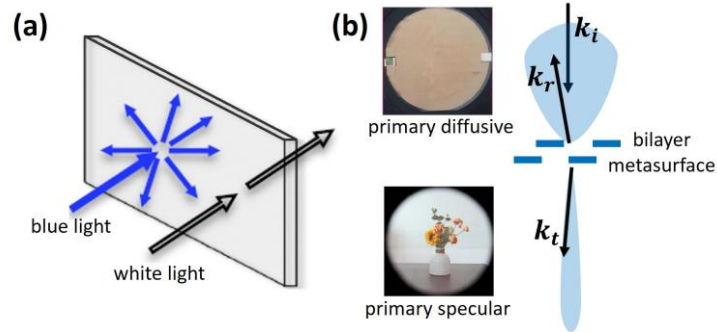

**Figure 12.** Two transparent display designs. **(a)** Form-factor engineering: metaatoms selectively resonant in the blue scatter blue light and preserve white light transmission. **(b)** Structure-factor engineering: bilayer metasurface offers completely different specular properties in reflection and transmission. The leftmost images are camera pictures of a flower bouquet taken in reflection (top) or transmission (bottom). Panel **(a)** reprinted with permission from Springer Nature: Hsu *et al.*, Nat. Commun. **5**, 3152 [169], Copyright 2014. Panel **(b)** from Chu *et al.*, Sci. Adv. **10**, eadm8061 (2024) [170]. Adapted with permission from AAAS.

## 8. Conclusion and perspectives

By nanostructuring surfaces with subwavelength features, notable changes can be induced in their scattering properties across both momentum and frequency domains. Over the past thirty years, there has been a notable transformation in the fundamental science driving this field of research. Advancements in scalable nanofabrication methods have precipitated a surge in the creation of nanostructured surfaces, now commonly referred to as 'optical metasurfaces'. In this broad landscape, disordered optical metasurfaces have only recently garnered significant attention. This new interest is not a result of lacking applications or curiosity. Rather, it is likely to be due to the lack of general electromagnetic modeling tools capable of adequately addressing the multiscale degrees of freedom inherent in these metasurfaces.

Therefore, rigorous advanced BSDF solvers which provide accurate predictions of the electromagnetic properties of disordered metasurfaces are pivotal for their development. However, they cannot stand on their own. To enable inverse design, we also need approximate semi-analytical models. The importance of developing approximate models that highlight the principal factors influencing the cross-link between nanostructure morphologies and both specular and diffuse light properties cannot be overstated.

Nanofabrication capabilities are rapidly advancing, receiving significant investments from the microelectronic industry and substantial attention from Tech Giants (Section 6). While the precision achieved through top-down approaches is undeniably attractive, the future of disordered metasurfaces likely resides in bottom-up approaches that are closer to meeting the requirements for industrial-scale production over large areas. Moreover, the ability to fabricate metasurfaces on non-flat surfaces with significant curvatures is another potential advantage of soft lithography and bottom-up methodologies in general [118,121,275].

Various metasurface morphologies can be fabricated with current technologies, yet only a few have captured research attention so far. These 'particular' metasurfaces consist of dilute monolayers of non-overlapping and distinct nanoparticles, for which advanced electromagnetic theories and modeling tools are starting to become broadly available.

For dilute monolayers of nanoparticles with a filling fraction of less than 10%, multiple scattering can often be neglected. Therefore, this review specifically emphasizes approximate models based on the Independent Scattering Approximation (ISA), see Section 5. In this context, we have demonstrated that two fundamental electromagnetic quantities, the form and structure factors, can be used to shape the BSDF.

For moderate density, near-field interactions and multiple scattering need to be considered. Mean-field theories are playing a crucial role, as they represent a natural expansion of the ISA. In Appendix C, we focus on approximate models that can predict the specular light scattered by infinite particular monolayers with greater accuracy than ISA models. Comparatively, fewer studies exist on diffuse light. Phenomenological correction terms can be incorporated into ISA models [276,79] to significantly increase predictive accuracy. It would be interesting to explore the possibility of obtaining accurate correction terms for the form factor using full-wave numerical methods.

At even higher densities, nanoparticles begin to overlap significantly, rendering traditional models ineffective. We can no longer consider the metasurfaces as particular monolayers; rather, we may consider them as two-phase nanostructures composed of various micro and nano morphologies. The properties of these metasurfaces remain largely unexplored [15].

There is still significant groundwork to be done in understanding and leveraging the relationship between morphology, material parameters, and the far-field scattering properties of disordered metasurfaces. We hope that the educational overview provided in Section 5 and Appendices A and B will help initiate further studies in this area.

Even with low-cost replication techniques, achieving large-scale and cost-effective production becomes challenging when precise control over the shape and size of the meta-atoms is required. Therefore, designing structures that are inherently resilient to material and shape variations is essential. An illustrative example is provided by nanoisland films on a dielectric spacer atop a metal substrate. Empirical observations indicate that the critical factor shaping the specular reflectance spectrum is the spacer thickness [173], and the spectrum is largely independent of the metal of the islands (refer to [17] for a plausible explanation). In this review, two additional examples in Section 6 have been analyzed, demonstrating how plasmonic resonances inherently robust against size and shape variations can be employed to achieve reproducible vivid colors despite polydispersity [85].

The application of disordered metasurfaces is still in its early stages, although many potential applications have existed for several decades. Section 7 provides an overview of recent applications, with a focus on areas where future research opportunities may arise.

Moreover, as modern integration technologies advance, there will be a growing demand for multifunctional electromagnetic devices. Consequently, one potential research direction could involve the development of cascaded disordered metasurfaces to integrate various functionalities, such as coloring, absorbing, or reflecting [277,278]. Another option lies in materials that enable deterministic, active tuning of metasurface responses, such as mesoporous [279], electrically-tunable [280] or microfluidic-driven [281,282] metasurfaces. These approaches could lead to paintings with changing appearances or colors.

Broadly speaking, progress in research on disordered structures and systems impacts not only optical metasurfaces but also other dimensions of photonics. The transverse localization of light in photonic systems incorporating disorder is another attractive topic. Explorations in this area have enhanced our understanding of fundamental optical phenomena, such as hyper-transport and localization with quantum-correlated photons [283]. A more comprehensive investigation of disordered metasurfaces may also benefit the design of disorder-immune systems, which are important in emerging areas of photonics, such as photonic quantum computing, topological or non-Hermitian photonics, and photonic machine learning [284].

After more than 30 years, research in disordered metasurfaces continues to progress steadily, suggesting that interest in this field is likely to endure. In summary, we hope this review stimulates further advancements at the intersection of disorder and metasurface-based optical devices, expanding the metasurface design toolbox with new approaches and possibilities.

## Appendix A: Specular reflection and transmission coefficients of infinite metasurfaces within the ISA

In this Note, using the ISA, we derive closed-form expressions for the reflection and transmission coefficients of infinite disordered metasurfaces with translation-invariant statistics. While the material is rather technical, the results obtained are relatively simple and intuitive. We begin by analyzing metasurfaces containing a finite number, N, of particles, before investigating the scenario where N tends to infinity while preserving a constant density.

We reintroduce the notations of the main text. We consider a collection of $N$ identical particles deposited on top of a stratified substrate. Particle $j$ is located at position $\mathbf{r}_j = [\mathbf{r}_{j,\parallel}, z_j = 0]$. The superstrate (upper hemisphere, labelled as '+') has a refractive index $n_{b,+}$ and the substrate bottom layer (lower hemisphere, labelled as '−') a refractive index $n_{b,-}$. The incident planewave has a frequency $\omega$ and the background field (in the absence of particles) is described by the electric field $\mathbf{E}_b(\mathbf{r}) = E_i\hat{\mathbf{e}}_i \exp[i\mathbf{k}_i \cdot \mathbf{r}] + r_{sub}E_i\hat{\mathbf{e}}_r \exp[i\mathbf{k}_r \cdot \mathbf{r}]$ in the upper hemisphere, with $E_i$ the incident planewave amplitude, $\hat{\mathbf{e}}_i$ and $\hat{\mathbf{e}}_r$ the unit polarization vectors of the incident and specularly reflected planewaves, and $\mathbf{k}_i = [\mathbf{k}_{i,\parallel}, k_{i,z}]$ and $\mathbf{k}_r = [\mathbf{k}_{i,\parallel}, -k_{i,z}]$ their wavevectors.

The reflection and transmission coefficients will be obtained, within the ISA, by summing the independent contributions of every particle in the far field. We thus start from the planewave decomposition of the field scattered by a single particle $j$. The decomposition requires the knowledge of two infinite sets (one for the superstrate '+' and the other '−' for the substrate) of $2 \times 2$ dyadic tensors. These tensors that are usually called Jones matrices, $\mathbf{J}(\mathbf{k}_{i,\parallel}, \mathbf{k}_{s,\parallel}, \pm)$, contain the scattering coefficients between the incident and scattered planewaves in the basis formed by two orthogonal polarizations (e.g., TE and TM). The set of matrices indicates how the scattered planewave amplitudes are distributed angularly, as illustrated in Figure 13. Typically, the matrices are computed using a Maxwell solver, followed by a near-to-far-field transformation (see Appendix C).

Note that Jones matrices can be thought as "transfer functions" that typically connect the excitation coefficients of plane waves that are normalized with a unitary Poynting vector. However, this convention is not adopted here. For simplicity, we will continue referring to them as Jones matrices.

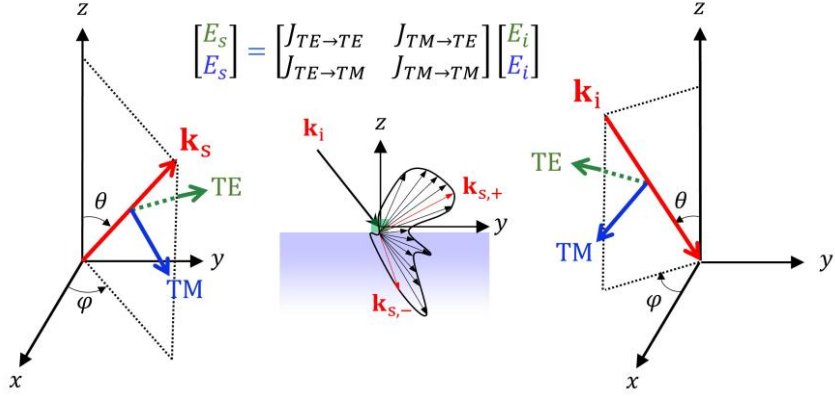

**Figure 13.** Illustration of the Jones matrix used for planewave expansions. The incident planewave (a 6-components distribution of the electric and magnetic fields at every point of the 3D space), is decomposed into a TE-polarized planewave (green) and a TM-polarized planewave (red). The same can be said between the scattered planewave. The scattered planewave is 'proportional' to the incident planewave, with a 'proportionality factor' that defines the 2 × 2 Jones matrix **J**. The latter can be computed by solving Maxwell equations for a single particle on the substrate and by applying a near-to-far-field transformation in the superstrate and substrate. It depends on many parameters, such as $\mathbf{k}_i$, $\mathbf{k}_s$, the particle itself, the layered substrate, the frequency.

Using Jones matrices, the planewave expansion of the field scattered by particle $j$ reads as

$$\mathbf{E}_s^{(j)}(\mathbf{r}) = \frac{E_i \exp[i\mathbf{k}_{i,\parallel}\cdot\mathbf{r}_{j,\parallel}]}{(2\pi)^2} \int \mathbf{J}\,\hat{\mathbf{e}}_i \exp\bigl[i\bigl(\mathbf{k}_\parallel \cdot (\mathbf{r}_\parallel - \mathbf{r}_{j,\parallel}) + \gamma_\pm|z|\bigr)\bigr]\, d\mathbf{k}_\parallel, \tag{A1}$$

where $\mathbf{J} \equiv \mathbf{J}(\mathbf{k}_{i,\parallel}, \mathbf{k}_\parallel, \pm)$, $\gamma_\pm = (k_{b,\pm}^2 - k_\parallel^2)^{1/2}$ with $\mathrm{Re}(\gamma_\pm) > 0$ and $\mathrm{Im}(\gamma_\pm) > 0$, $k_{b,\pm} = n_{b,\pm}\omega/c$. Note that the integral in Equation A1 is performed on the entire 2D reciprocal space, thereby comprising both propagating and evanescent waves. We may rewrite Equation A1 as

$$\mathbf{E}_s^{(j)}(\mathbf{r}) = \frac{E_i}{(2\pi)^2} \int \mathbf{J}\,\hat{\mathbf{e}}_i \exp\bigl[-i(\mathbf{k}_\parallel - \mathbf{k}_{i,\parallel}) \cdot \mathbf{r}_{j,\parallel}\bigr] \exp[i(\mathbf{k}_\parallel \cdot \mathbf{r}_\parallel + \gamma_\pm|z|)]\, d\mathbf{k}_\parallel. \tag{A2}$$

Within the ISA, every individual particle is excited by the same incident field, up to a phase factor $\exp[i\mathbf{k}_{i,\parallel}\cdot\mathbf{r}_{j,\parallel}]$. The scattered field from the collection of particles is thus given by

$$\sum_{j=1}^N \mathbf{E}_s^{(j)}(\mathbf{r}) = \frac{E_i}{(2\pi)^2} \int \mathbf{J}\,\hat{\mathbf{e}}_i \left(\sum_{j=1}^N \exp[-i(\mathbf{k}_\parallel - \mathbf{k}_{i,\parallel}) \cdot \mathbf{r}_{j,\parallel}]\right) \exp[i(\mathbf{k}_\parallel \cdot \mathbf{r}_\parallel + \gamma_\pm|z|)]\, d\mathbf{k}_\parallel. \tag{A3}$$

Taking an average over disorder realizations (i.e., on the statistical positions of the particles), we get

$$\left\langle\sum_{j=1}^N \mathbf{E}_s^{(j)}(\mathbf{r})\right\rangle = \frac{E_i}{(2\pi)^2} \int \mathbf{J}\,\hat{\mathbf{e}}_i \left\langle\sum_{j=1}^N \exp[-i(\mathbf{k}_\parallel - \mathbf{k}_{i,\parallel}) \cdot \mathbf{r}_{j,\parallel}]\right\rangle \exp[i(\mathbf{k}_\parallel \cdot \mathbf{r}_\parallel + \gamma_\pm|z|)]\, d\mathbf{k}_\parallel. \tag{A4}$$

The ensemble average is defined as $\langle f(\mathbf{R})\rangle = \int f(\mathbf{R}) P(\mathbf{R}) d\mathbf{R}$, where $P(\mathbf{R})$ represents the probability density function for specific disorder realizations and $d\mathbf{R} = d\mathbf{r}_1 d\mathbf{r}_2 \ldots d\mathbf{r}_N$. We assume that all particles are independently positioned on a surface $\Sigma$ and follow identical statistical distributions. We have $P(\mathbf{R}) = P(\mathbf{r}_1) P(\mathbf{r}_2) \ldots P(\mathbf{r}_N)$ with $P(\mathbf{r}_j) = \Sigma^{-1}$. Taking the limit of an infinite size with a constant surface density $\lim_{N,\Sigma\to\infty} \frac{N}{\Sigma} = \rho$, we identify the 2D Fourier transform of the constant function and obtain

$$\lim_{N,\Sigma\to\infty} \left\langle\sum_{j=1}^N \exp[-i(\mathbf{k}_\parallel - \mathbf{k}_{i,\parallel}) \cdot \mathbf{r}_{j,\parallel}]\right\rangle = \rho\,(2\pi)^2 \delta(\mathbf{k}_\parallel - \mathbf{k}_{i,\parallel}), \tag{A5}$$

where $\delta(\mathbf{k})$ is the Dirac delta distribution. Equation A4 leads to

$$\lim_{N,\Sigma \to \infty} \left\langle \sum_{j=1}^N \mathbf{E}_s^{(j)}(\mathbf{r}_\parallel, z) \right\rangle = E_i \rho \, \mathbf{J}(\mathbf{k}_{i,\parallel}, \mathbf{k}_{i,\parallel}, \pm) \, \hat{\mathbf{e}}_i \exp[i(\mathbf{k}_{i,\parallel} \cdot \mathbf{r}_\parallel + \gamma_{i,\pm}|z|)], \tag{A6}$$

where $\gamma_{i,\pm} = \sqrt{k_{b,\pm}^2 - k_{i,\parallel}^2}$ with $\text{Re}(\gamma_{i,\pm}) > 0$ and $\text{Im}(\gamma_{i,\pm}) > 0$. It is worth noting that we have reintroduced the explicit dependence of $\mathbf{J}(\mathbf{k}_{i,\parallel}, \mathbf{k}_{i,\parallel}, \pm)$ in Equation A6 to accommodate the Dirac distribution. The Dirac distribution indicates that the average scattered field in the scenario of infinite disordered metasurfaces with translation-invariant statistics is a planewave with a parallel wavevector equal to the incident planewave, $\mathbf{k}_{i,\parallel}$. Consequently, two solutions are plausible, corresponding to specular reflection and transmission.

In reflection ($z > 0$), we obtain the specular reflected planewave

$$\langle \mathbf{E}_r(\mathbf{r}) \rangle_\infty = r_{\text{sub}} E_i \hat{\mathbf{e}}_r \exp[i\mathbf{k}_r \cdot \mathbf{r}] + \lim_{N,\Sigma \to \infty} \left\langle \sum_{j=1}^N \mathbf{E}_s^{(j)}(\mathbf{r}_\parallel, z) \right\rangle$$
$$= E_i [r_{\text{sub}} \hat{\mathbf{e}}_r + \rho \, \mathbf{J}(\mathbf{k}_{i,\parallel}, \mathbf{k}_{i,\parallel}, +) \, \hat{\mathbf{e}}_i] \exp[i\mathbf{k}_r \cdot \mathbf{r}]. \tag{A7}$$

Note the subscript "$\infty$" that we have introduce in the statistical averaging $\langle \cdot \rangle_\infty$. It indicates that the result is obtained for an infinite metasurface.

In transmission ($z < 0$), defining $\mathbf{k}_t = \left[\mathbf{k}_{i,\parallel}, \sqrt{k_{b,-}^2 - k_{i,\parallel}^2}\right]$ as the wavevector of the transmitted planewave and $\hat{\mathbf{e}}_t$ its unit polarization vector, we obtain for the specular transmitted planewave

$$\langle \mathbf{E}_t(\mathbf{r}) \rangle_\infty = t_{\text{sub}} E_i \hat{\mathbf{e}}_t \exp[i\mathbf{k}_t \cdot \mathbf{r}] + \lim_{N,\Sigma \to \infty} \left\langle \sum_{j=1}^N \mathbf{E}_s^{(j)}(\mathbf{r}_\parallel, z) \right\rangle$$
$$= E_i [t_{\text{sub}} \hat{\mathbf{e}}_t + \rho \, \mathbf{J}(\mathbf{k}_{i,\parallel}, \mathbf{k}_{i,\parallel}, -) \, \hat{\mathbf{e}}_i] \exp[i\mathbf{k}_t \cdot \mathbf{r}]. \tag{A8}$$

Finally, we get the reflection coefficient of the monolayer for a polarization described by $\hat{\mathbf{e}}_r$ is given by $r_{ISA} E_i \exp[i\mathbf{k}_r \cdot \mathbf{r}] = \hat{\mathbf{e}}_r \cdot \langle \mathbf{E}_r(\mathbf{r}) \rangle$, leading to

$$r_{ISA} = r_{\text{sub}} + \rho \left( \hat{\mathbf{e}}_r \cdot \mathbf{J}(\mathbf{k}_{i,\parallel}, \mathbf{k}_{i,\parallel}, +) \, \hat{\mathbf{e}}_i \right). \tag{A9}$$

Similarly, the transmission coefficient of the monolayer is given by $t_{ISA} E_i \exp[i\mathbf{k}_t \cdot \mathbf{r}] = \hat{\mathbf{e}}_t \cdot \langle \mathbf{E}_t(\mathbf{r}) \rangle$, leading to

$$t_{ISA} = t_{\text{sub}} + \rho \left( \hat{\mathbf{e}}_t \cdot \mathbf{J}(\mathbf{k}_{i,\parallel}, \mathbf{k}_{i,\parallel}, -) \, \hat{\mathbf{e}}_i \right). \tag{A10}$$

Due to interference effects, the coefficients $r$ and $t$ may either exceed or fall below the values of $r_{\text{sub}}$ and $t_{\text{sub}}$ for the layered substrate devoid of particles.

### Appendix B: Far-field diffuse intensity of infinite metasurfaces within the ISA

In this note, we utilize the ISA to establish analytical expressions for the far-field diffuse intensity of infinite disordered metasurfaces characterized by translation-invariant statistics. Despite the technical depth of the content, the resulting conclusions are straightforward. As with Appendix A, we begin by examining metasurfaces containing a finite number of particles before extending our analysis to scenarios where the particle count approaches infinity while maintaining a constant density.

**Finite metasurfaces.** For the derivation, it is essential to recognize that the angle-dependent intensity (expressed in reciprocal space) is the 2D Fourier transform of the field-field correlation $\langle \mathbf{E}_s(\mathbf{r}) \cdot \mathbf{E}_s^\star(\mathbf{r}') \rangle$. Let us then start by writing the correlation of the fields created by two particles $j$ and $l$ in a plane $z$ above or below the monolayer. From Equation A1), we get

$$\mathbf{E}_s^{(j)}(\mathbf{r}_\parallel, z) \cdot \mathbf{E}_s^{(l)\star}(\mathbf{r}_\parallel', z) = |E_i|^2 \int \left( [\mathbf{J}(\mathbf{k}_{i,\parallel}, \mathbf{k}_\parallel, \pm) \, \hat{\mathbf{e}}_i] \cdot [\mathbf{J}^\star(\mathbf{k}_{i,\parallel}, \mathbf{k}_\parallel', \pm) \, \hat{\mathbf{e}}_i] \right)$$
$$\times \exp[i(\mathbf{k}_\parallel \cdot (\mathbf{r}_\parallel - \mathbf{r}_{j,\parallel}) + \gamma_\pm |z|)] \exp[-i(\mathbf{k}_\parallel' \cdot (\mathbf{r}_\parallel' - \mathbf{r}_{l,\parallel}) + \gamma_\pm'^\star |z|)]$$

$$\times \exp[i\mathbf{k}_{i,\parallel} \cdot (\mathbf{r}_{j,\parallel} - \mathbf{r}_{l,\parallel})] \frac{d\mathbf{k}_\parallel}{(2\pi)^2} \frac{d\mathbf{k}'_\parallel}{(2\pi)^2}. \tag{B1}$$

To proceed, we make the following changes of variables: $\mathbf{k}_\parallel = \mathbf{K} + \mathbf{Q}/2$, $\mathbf{k}'_\parallel = \mathbf{K} - \mathbf{Q}/2$, $\mathbf{r}_\parallel = \mathbf{R} + \mathbf{X}/2$, and $\mathbf{r}_\parallel' = \mathbf{R} - \mathbf{X}/2$. Equation B1 becomes (we drop $\mathbf{k}_{i,\parallel}$ and $\pm$ in $\mathbf{J}$ and $\mathbf{J}^\star$ for simplicity)

$$\mathbf{E}_s^{(j)}\left(\mathbf{R} + \frac{\mathbf{X}}{2}, z\right) \cdot \mathbf{E}_s^{(l)\star}\left(\mathbf{R} - \frac{\mathbf{X}}{2}, z\right) = |E_i|^2 \int ([\mathbf{J}(\mathbf{K} + \mathbf{Q}/2)\,\hat{\mathbf{e}}_i] \cdot [\mathbf{J}^\star(\mathbf{K} - \mathbf{Q}/2)\,\hat{\mathbf{e}}_i])$$

$$\times \exp[i\mathbf{K} \cdot \mathbf{X}]\exp[i\mathbf{K} \cdot (\mathbf{r}_{l,\parallel} - \mathbf{r}_{j,\parallel})]\exp[i\mathbf{Q} \cdot \mathbf{R}]\exp[-i\mathbf{Q} \cdot (\mathbf{r}_{j,\parallel} + \mathbf{r}_{l,\parallel})/2]$$

$$\times \exp[i(\gamma_\pm - \gamma'^\star_\pm)|z|]\exp[i\mathbf{k}_{i,\parallel} \cdot (\mathbf{r}_{j,\parallel} - \mathbf{r}_{l,\parallel})] \frac{d\mathbf{K}}{(2\pi)^2} \frac{d\mathbf{Q}}{(2\pi)^2}. \tag{B2}$$

Within the ISA, the field-field correlation from an assembly of identical (and independent) particles averaged over disorder realizations is thus

$$\left\langle \sum_{j,l=1}^N \mathbf{E}_s^{(j)}\left(\mathbf{R} + \frac{\mathbf{X}}{2}, z\right) \cdot \mathbf{E}_s^{(l)\star}\left(\mathbf{R} - \frac{\mathbf{X}}{2}, z\right) \right\rangle = |E_i|^2 \int \left([\mathbf{J}\left(\mathbf{K} + \frac{\mathbf{Q}}{2}\right)\hat{\mathbf{e}}_i] \cdot [\mathbf{J}^\star\left(\mathbf{K} - \frac{\mathbf{Q}}{2}\right)\hat{\mathbf{e}}_i]\right)$$

$$\times \left\langle \sum_{j,l=1}^N \exp[-i(\mathbf{K} - \mathbf{k}_{I,\parallel}) \cdot (\mathbf{r}_{j,\parallel} - \mathbf{r}_{l,\parallel})]\exp[-i\mathbf{Q} \cdot (\mathbf{r}_{j,\parallel} + \mathbf{r}_{l,\parallel})/2] \right\rangle$$

$$\times \exp[i\mathbf{K} \cdot \mathbf{X}]\exp[i\mathbf{Q} \cdot \mathbf{R}]\exp[i(\gamma_\pm - \gamma'^\star_\pm)|z|] \frac{d\mathbf{K}}{(2\pi)^2} \frac{d\mathbf{Q}}{(2\pi)^2}. \tag{B3}$$

We further proceed by calculating taking a 2D spatial average of the field-field correlation over $\mathbf{R}$ (which is nothing but the mid-point between pairs of particles, $\mathbf{R} = (\mathbf{r}_\parallel + \mathbf{r}'_\parallel)/2$). We define the average field-field correlation as $\mathcal{L}(\mathbf{X}) \equiv \int \left\langle \sum_{j,l=1}^N \mathbf{E}_s^{(j)}(\mathbf{R} + \mathbf{X}/2, z) \cdot \mathbf{E}_s^{(l)\star}(\mathbf{R} - \mathbf{X}/2, z) \right\rangle d\mathbf{R}$. In Eq. (B3), the sole reliance on $\mathbf{R}$ occurs within the exponential term $\exp[i\mathbf{Q} \cdot \mathbf{R}]$. Utilizing $\int \exp[i\mathbf{Q} \cdot \mathbf{R}]\,d\mathbf{R} = (2\pi)^2 \delta(\mathbf{Q})$, we integrate over $\mathbf{Q}$ and obtain

$$\mathcal{L}(\mathbf{X}) = |E_i|^2 \int \|\mathbf{J}(\mathbf{K})\,\hat{\mathbf{e}}_i\|^2 \left\langle \sum_{j,l=1}^N \exp[i(\mathbf{K} - \mathbf{k}_{i,\parallel}) \cdot (\mathbf{r}_{l,\parallel} - \mathbf{r}_{j,\parallel})] \right\rangle$$

$$\times \exp[i\mathbf{K} \cdot \mathbf{X}]\exp[-2\mathrm{Im}(\gamma_\pm)|z|] \frac{d\mathbf{K}}{(2\pi)^2}. \tag{B4}$$

Note that $\mathbf{K} = (\mathbf{k}_\parallel + \mathbf{k}'_\parallel)/2$ is the average in-plane wavevector between $\mathbf{k}_\parallel$ and $\mathbf{k}'_\parallel$, and the integral over $\mathbf{K}$ is performed over the entire reciprocal space, including evanescent and propagating planewaves. The angle-resolved far-field scattered intensity is therefore obtained by evaluating the 2D Fourier transform at $\mathbf{K} = \mathbf{k}_{s,\parallel}$,

$$\mathcal{L}(\mathbf{k}_s, \mathbf{k}_i) = |E_i|^2 \|\mathbf{J}(\mathbf{K})\,\hat{\mathbf{e}}_i\|^2 \left\langle \sum_{j,l=1}^N \exp[-i(\mathbf{k}_{s,\parallel} - \mathbf{k}_{i,\parallel}) \cdot (\mathbf{r}_{j,\parallel} - \mathbf{r}_{l,\parallel})] \right\rangle. \tag{B5}$$

We can finally rewrite Equation B5 for the far-field scattered intensity $\mathcal{L}$ as

$$\mathcal{L}(\mathbf{k}_s, \mathbf{k}_i) = N\,|E_i|^2\,F(\mathbf{k}_{s,\parallel}, \pm, \mathbf{k}_{i,\parallel}, \hat{\mathbf{e}}_i)\,S(\mathbf{k}_{s,\parallel} - \mathbf{k}_{i,\parallel}), \tag{B6}$$

where we have introduced the *form factor* (we bring back the explicit dependence of $\mathbf{J}$ for this final result)

$$F(\mathbf{k}_{s,\parallel}, \pm, \mathbf{k}_{i,\parallel}, \hat{\mathbf{e}}_i) = \|\mathbf{J}(\mathbf{k}_{i,\parallel}, \mathbf{k}_{s,\parallel}, \pm)\,\hat{\mathbf{e}}_i\|^2, \tag{B7}$$

and the *static* structure factor (also introduced in Equation 11)

$$S(\mathbf{q}) = \frac{1}{N}\left\langle \sum_{j,l=1}^N \exp[-i\mathbf{q} \cdot (\mathbf{r}_{j,\parallel} - \mathbf{r}_{l,\parallel})] \right\rangle. \tag{B8}$$

In Appendix C, we illustrate how the form factor can be computed with near-to-far-field transforms.

**Infinite metasurfaces.** We now proceed to the limit of infinite metasurfaces for asymptotically large values of $N$ and for a fixed density. The following calculations of the structure factor with the introduction of the pair correlation function are classical.

We begin by conveniently rewriting the static structure factor as the sum of two terms, one for $j = l$ and the other for $j \neq l$, $S(\mathbf{q}) = 1 + \frac{1}{N}\left\langle \sum_{j \neq l, 1}^N \exp[-i\mathbf{q} \cdot (\mathbf{r}_{j,\parallel} - \mathbf{r}_{l,\parallel})] \right\rangle$. Assuming that all particle pairs are *identically and independently distributed*, we obtain

$$S(\mathbf{k}_{s,\|} - \mathbf{k}_{i,\|}) = 1 + \frac{1}{N}N(N-1)\langle\exp[-i(\mathbf{k}_{s,\|} - \mathbf{k}_{i,\|})\cdot(\mathbf{r}_{1,\|} - \mathbf{r}_{2,\|})]\rangle .$$ The ensemble average of the last term can be written explicitly as

$$\langle\exp[-i(\mathbf{k}_{s,\|} - \mathbf{k}_{i,\|})\cdot(\mathbf{r}_{1,\|} - \mathbf{r}_{2,\|})]\rangle = \int \exp[-i(\mathbf{k}_{s,\|} - \mathbf{k}_{i,\|})\cdot(\mathbf{r}_{1,\|} - \mathbf{r}_{2,\|})] P(\mathbf{r}_{1,\|}, \mathbf{r}_{2,\|})\mathrm{d}\mathbf{r}_{1,\|}\mathrm{d}\mathbf{r}_{2,\|}. \tag{B9}$$

By introducing the joint probability to find a particle at position $\mathbf{r}_{1,\|}$ and another particle at position $\mathbf{r}_{2,\|}$, $P^{(2)}(\mathbf{r}_{1,\|}, \mathbf{r}_{2,\|}) = N(N-1)P(\mathbf{r}_{1,\|}, \mathbf{r}_{2,\|})$, we obtain

$$S(\mathbf{k}_{s,\|} - \mathbf{k}_{i,\|}) = 1 + \frac{1}{N}\int \exp[-i(\mathbf{k}_{s,\|} - \mathbf{k}_{i,\|})\cdot(\mathbf{r}_{1,\|} - \mathbf{r}_{2,\|})] P^{(2)}(\mathbf{r}_{1,\|}, \mathbf{r}_{2,\|})\mathrm{d}\mathbf{r}_{1,\|}\mathrm{d}\mathbf{r}_{2,\|}. \tag{B10}$$

When we additionally substitute the variable $\mathbf{r}_{2,\|}$ with $\mathbf{r}_{1,\|} - \mathbf{P}$, take the limit of $N, \Sigma \to \infty$ with $\lim_{N,\Sigma\to\infty}\frac{N}{\Sigma} = \rho$, and assume that the medium is statistically translationally-invariant, $P^{(2)}(\mathbf{r}_{1,\|}, \mathbf{r}_{2,\|}) = P^{(2)}(\mathbf{r}_{1,\|} - \mathbf{r}_{2,\|}) = P^{(2)}(\mathbf{P})$. Then, the integral over $\mathrm{d}\mathbf{r}_{1,\|}$ gives $\Sigma$ and we obtain

$$\lim_{N,\Sigma\to\infty} S(\mathbf{k}_{s,\|} - \mathbf{k}_{i,\|}) = 1 + \frac{1}{\rho}\int \exp[-i(\mathbf{k}_{s,\|} - \mathbf{k}_{i,\|})\cdot\mathbf{P}] P^{(2)}(\mathbf{P})\mathrm{d}\mathbf{P}. \tag{B11}$$

Finally, introducing the pair correlation function $g_2(\mathbf{r}) = P^{(2)}(\mathbf{r})/\rho^2$, we obtain

$$\lim_{N,\Sigma\to\infty} S(\mathbf{k}_{s,\|} - \mathbf{k}_{i,\|}) = 1 + \rho \int \exp[-i(\mathbf{k}_{s,\|} - \mathbf{k}_{i,\|})\cdot\mathbf{P}] g_2(\mathbf{P})\mathrm{d}\mathbf{P}. \tag{B12}$$

## Appendix C. Computation and use of Jones matrices $\mathbf{J}(\mathbf{k}_{i,\|}, \mathbf{k}_{s,\|}, \pm)$

In this note, we provide an easy-to-use guideline to compute:

- the Jones matrix of a single metaatom of any shape,
- the specular reflection and transmission coefficients of disordered monolayers on substrate.

We assume that the layered substrate consists of isotropic materials.

*Jones matrix computation*

The Jones matrix of a single metaatom of any shape can be computed by using near-to-far-field transformation. We explain how to perform this computation by using the RETOP toolbox [87] developed in the group of the first author. These guidelines may also be useful for other software.

The first step in the near-to-far-field transformation is the pre-calculation of the near-zone field $(\mathbf{E}, \mathbf{H})$ scattered by the particle under illumination by the incident planewave $E_i \exp[i\mathbf{k}_{i,\|}\cdot\mathbf{r}_{j,\|}]\hat{\mathbf{e}}_i$. Two independent computations are performed for TE ($\hat{\mathbf{e}}_i$ perpendicular to the incidence plane) and TM ($\hat{\mathbf{e}}_i$ in the incidence plane) polarizations. The near-zone fields should be computed on a closed box surrounding the particle, positioned in its environment including the layered substrate. This computation can be performed using any Maxwell solver. They are then used by RETOP to calculate two radiation diagrams in the substrate and the superstrate.

The **retop.m** function generates a Matlab structure named angles. The Jones matrices are derived from two outputs: angles_up.EE or angles_dn.EE, which correspond to the radiation diagrams in the superstrate '+' and substrate '−', respectively. The 1×2 array [angles_up.EE(:,1), angles_up.EE(:,2)] represents the electric-field components of the planewaves propagating in the $\mathbf{k}_s$ direction in the (TM,TE) basis in the superstrate (Figure 13). Similarly [angles_dn.EE(:,1), angles_dn.EE(:,2)] represents the electric-field components of the planewaves propagating in the $\mathbf{k}_s$ direction in the (TM,TE) basis in the substrate.

As computations are necessary for both TE and TM incident polarizations, we obtain four outputs for the substrate and four other outputs for the superstrate. Every four outputs correspond to the four Jones coefficients, as illustrated in Figure 13, after normalization by the incident electric field.

Any plane wave expansion has a phase origin, which must be controlled to accurately compute specular reflection and transmission, as described below. We recommend setting the origin at the plane of the interface where the particles are deposited. In RETOP, the phase origins in the superstrate and substrate are given by two real numbers: angles_up.origine and angles_dn.origine, which can be used to adjust the phase origin to the interface.

*Computation of the specular reflection and transmission coefficients*

We use MATLAB language to compute the specular reflection and transmission coefficients. For these coefficients, we just need the Jones matrix $\mathbf{J}(\mathbf{k}_{i,\parallel}, \mathbf{k}_{s,\parallel}, \pm)$ with $\mathbf{k}_{s,\parallel} = \mathbf{k}_{i,\parallel}$. We first need to identify which column element of the angles_up.EE or angles_dn.EE arrays corresponds to the specular direction. We denote its rank by 'number' and we simplify notations:

>> EEh = angles_up.EE(number,:); EEb = angles_dn.EE(number,:); % EEh and EEb are 1×2 arrays

In order to calculate the Jones matrix of the isolated metaatom, we need to know the electric field amplitude of the input planewave at the phase origin. This is obtained from the Maxwell equation software (not RETOP) used to compute the scattered field on the box surrounding the particle. We may then define a 1×2 vector

>> EEinc = [EEincTM, EEincTE]; % EEincTM (resp. EEincTE) being the electric field amplitude $E_i$ of the incident planewave for TM (resp. TE) polarization

Then we compute the Jones matrix $\mathbf{J}(\mathbf{k}_{i,\parallel}, \mathbf{k}_{s,\parallel}, \pm)$ with $\mathbf{k}_{s,\parallel} = \mathbf{k}_{i,\parallel}$. In general, we may have polarization conversion even in the specular direction. To simplify, we assume that the monolayers have statistically invariant distribution of nanoparticles and that the nanoparticles are rotationally invariant. In this case the Jones matrix is a diagonal matrix, and we obtain

>> % TE polarization
>> Jh = 4*pi*EEh(2)/EEinc(2); Jb = 4*pi*EEb(2)/EEinc(2);
>> % TM polarization
>> Jh = 4*pi*EEh(1)/EEinc(1); Jb = 4*pi*EEb(1)/EEinc(1);

The Jones matrix in the specular reflection and transmission direction can be injected into Equations 16a and 16b (ISA model),

% Single scattering model with substrate (Equations 16a and 16b)
>> r_stack_ISA = r_sub + rho*Jh; % rho is the density & r_sub is the substrate reflection coefficient without nanoparticle
>> t_stack_ISA = t_sub + rho *Jb; % t_sub is the substrate transmission coefficient without nanoparticle

To achieve better accuracy, we may use the EFA model (Equations 18a and 18b) or, even better, the Quasi-Crystalline Approximation (QCA) model [109] to compute the specular reflection and transmission coefficients of infinite monolayers of disordered metaatoms. For that purpose, the Jones matrix computation should be performed in a uniform background with a refractive index n_b, corresponding to the medium surrounding the particles (air in the example of Figure 14). The ISA, EFA and QCA models give us

% Independent scattering approximation (ISA) model
>> r_ISA = rho *Jh; % rho is the density
>> t_ISA = 1 + rho *Jb;

% Effective-field (EFA) model (Equations 18a and 18b)
>> r_EFA = rho *Jh./ (1 - rho *Jb);
>> t_EFA = 1 ./ (1 - rho *Jb);

% Quasi-Crystalline Approximation (QCA) model [Garcia-Valenzuela et al, JOSA A 29, 1161 (2012)]
>> r_QCA = rho *Jh./ (1 - dens* Jb + rho ^2/4 * (Jb.^2-Jh.^2));

```
>> t_QCA = (1 - rho ^2/4 * (Jb.^2- Jh.^2)) ./ (1 + rho * Jb + rho ^2/4 * (Jb.^2-Jh.^2));
```

These formulas are valid only in uniform backgrounds. Once the reflection and transmission coefficients for the monolayer in the uniform background are obtained with one of the three models, the impact of a substrate on specular reflectance and transmittance is accounted for via Fresnel equations.

We define first r_coh and t_coh

```
>> r_coh = r_ISA; t_coh = t_ISA ; % for the ISA model
>> r_coh = r_EFA; t_coh = t_EFA ; % for the EFA model
>> r_coh = r_QCA; t_coh = t_QCA ; % for the QCA model
```

The reflection coefficients r_stack and t_stack of a monolayer above a layered substrate are then given by

```
>> r_stack = r_coh + (r_sub * t_coh^2 *uu^2) / (1 - r_sph * r_coh *uu^2);
>> t_stack = sqrt(n_sub*cos(asin(n_sup/n_sub*sin(theta_inc)))/(n_sup*cos(theta_inc))) * t_stack_n;
```

with

```
>> uu = exp(1i*k0*r_sph*n_b*cos(theta))
>> t_stack_n = (t_sub * t_coh *uu) / (1 - r_sub * r_coh *uu^2);
```

In these expressions, r_coh and t_coh are the reflection and transmission coefficients of the particular monolayer – determined with any preferred model. n_b is the refractive index of the medium in which the particles are embedded, n_sub is the refractive index of the substrate, n_sup is the refractive index of the superstrate, and $r\_sph$ is the height separating the monolayer and the first substrate interface (it is also the sphere radius). r_sub and t_sub can be calculated using, for instance, the 2 × 2 transfer matrix method. k0 is the modulous of the wavevector in vacuum. theta_inc is the angle of incidence in the superstrate. theta is such that n_sup*sin(theta_inc) = n_b*sin(theta).

Figure 14 shows the specular reflection and transmission spectra of a monolayer of disordered Si nano-sphere (radius 50 nm, 10% filling fraction) for an incident planewave angle of 60° and TM polarization. One can see the spectra exhibit two sharp peaks due to the Mie resonances of the individual particles. The strength of these resonances is yet strongly overestimated by the ISA model compared to EFA and QCA models. The ISA model also leads to unphysical result, i.e., transmission larger than 1 for $\lambda \approx 500$ nm. On request, the first author may provide the program used for the simulations. Other numerical results and comparisons with full-wave numerical data for related geometries can be found in [96].

## Appendix D. Classical static structure factors

For randomly positioned particles, the $\mathbf{r}_{j,\parallel}$'s are independent variables and the static structure factor (Equation 14) of infinite metasurfaces is independent of $\mathbf{q}_{\parallel}$ ($S_{r,\infty} = 1$). To implement pronounced spectral and angular features in the diffuse light, one should implement structural correlations.

The importance of structural correlations to harness light transport in disordered systems is not a new topic. Research on correlated disordered media in optics however really took off in the mid-2000s with experimental studies showing that disorder could be engineered to harness light transport in 3D materials. Section 7 illustrates a few interesting applications using structural correlations, not to harness light transport through a medium but to control the spectral and angular spectra of the field scattered by 2D metasurfaces, for instance for absorption [22,162,244,248,249], extraction [257] or transparency [164].

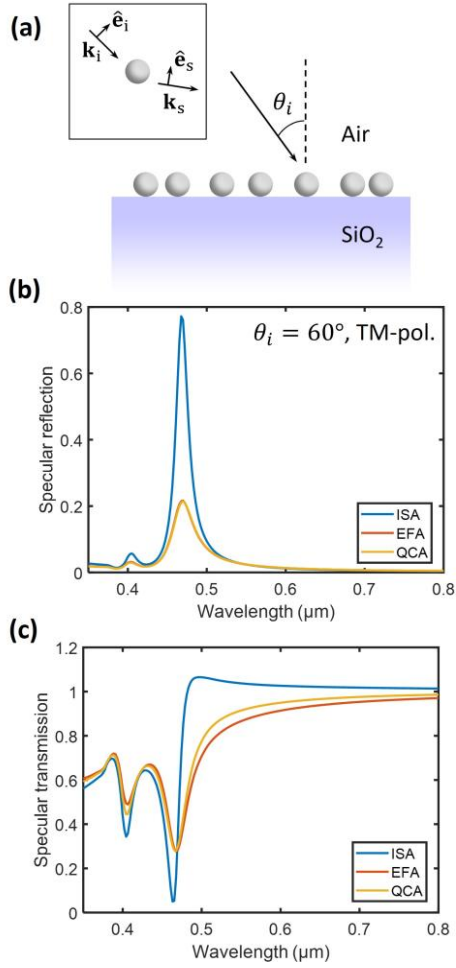

**Figure 14.** (a) The Schematic of a monolayer of 50-nm-radius Silicon nanoparticles (sphere) deposited at a surface coverage (or filling fraction) $f = 0.10$ on **top** of a $SiO_2$ substrate. Inset illustrates the scattering by one silicon sphere in a uniform medium. **(b)-(c)** Specular reflection and transmission spectra of the structure, with incident angles 60°, TM polarization.

Most applications rely on a few common static structure factors outlined in Figure 15. For the sake of clarity, we provide 2D maps $S(q_x, q_y)$, that remain unchanged under rotation and a radial plot that quantitatively depicts $S(q)$. Figure 15a addresses short-range amorphous-type correlations, distinguished by a hard exclusion disk of diameter $a$. These disks, depicted by grey circles, ensures that the interparticle distance between all the particles of the random array is larger than $a$ [285]. Such correlated disorders are controlled by the packing fraction, $p = \rho \pi a^2/4$, where $\rho$ again denotes the particle density. Increasing density or minimum interparticle distance introduces short-range correlations, leading to a decrease in the structure factor at low $q$ values and the appearance of a peak just below $qa = 2\pi$. The reduction of the diffuse light around the specular direction $\mathbf{q}_{||} = \mathbf{0}$ has been observed experimentally for normal and oblique incidences [23,79] and has been explored further in Section 7 for device design purposes.

Figure 15b depicts a distinctly different disorder associated with hyperuniformity, a concept tied to spatial regularity, which quantifies how the number of points within a given area varies across different realizations of disorder. In hyperuniform structures, a combination of short- and long-range interparticle interactions leads to the static structure factor approaching zero as $q$ approaches zero [246,286]. Figure 15b refers to a specific hyperuniformity, known as "stealthy" hyperuniformity, where $S_r(q) = 0$ for $|q| < q_{max}$, with $q_{max}$ representing an arbitrary threshold.

Today, it is feasible to generate extensive sets of points embodying correlated disorders using various freely available software programs [91-95]. Importantly, this generation process is not confined to statistically isotropic ensembles; one can explore anisotropic hyperuniformity, wherein distinct responses for $q_x$ and $q_y$ are considered [18].

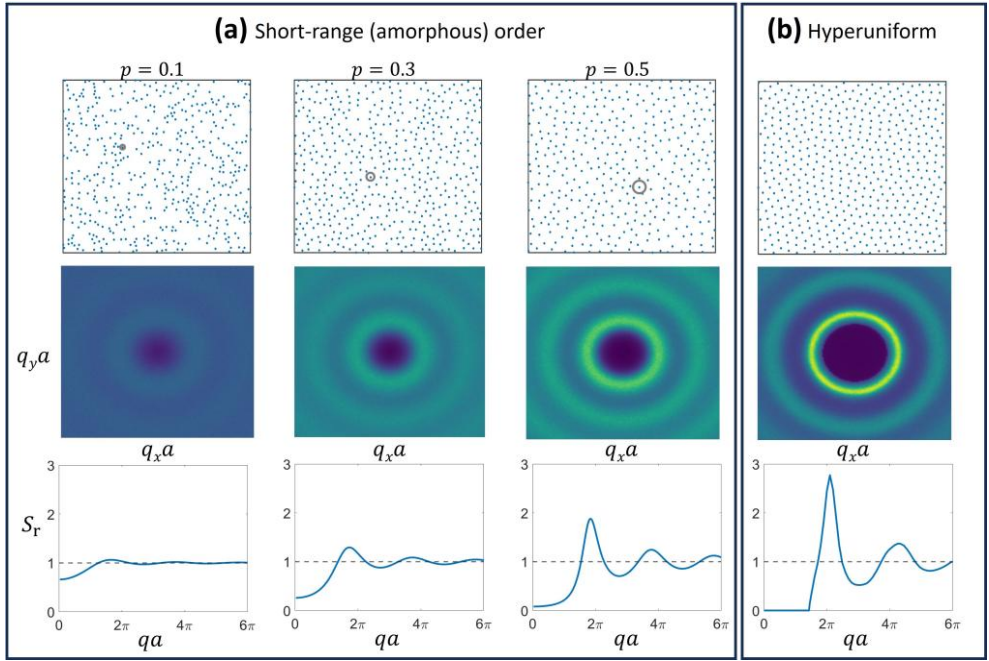

**Figure 15.** Structure factors of a few classical engineered disorders. $\boldsymbol{q} = [q_x, q_y, 0]^T$ signifies the difference in wavevector between the scattered and incident directions and $q = |\boldsymbol{q}|$. **(a)** Short-range amorphous-type disorder. The static structure factor is computed from the semi-analytical Baus-Colot model [285] and $a$ represents the imposed minimum particle (center-to-center) separation distance. **(b)** Structure factor of a "stealthy" hyperuniform structures, for which $S_r(q) = 0$ for $|q| < q_{max}$, with $q_{max}$ an arbitrary value. In this case, $a$ is defined by $\rho a^2 = 1$.

**Acknowledgments.**

AD acknowledges the Swedish Research Council for Sustainable Development (Formas) (Project No. 2021-01390), Thuréus Forskarhem och Naturminne Foundation, and the Swedish Research Council (VR) (Project No. 2024-05025). PL and KV acknowledge financial support from the French National Agency for Research (ANR) under the project "NANO-APPEARANCE" (ANR-19-CE09-0014). PL acknowledges financial support from the European Research Council Advanced grant (Project UNSEEN No. 101097856). CR and AS acknowledge financial support from the Deutsche Forschungsgemeinschaft (DFG) (Nos. 413644979, RO 3640/11-1, and WE 4051/26-1). The authors wish to thank fruitful interactions with Romain Pacanowski, Pascal Barla, Xavier Granier, Franck Carcenac, Glenna Drisko, Mona Treguet-Delapierre, Armel Pitelet, and Louis Bellando.

**Disclosures.**

The authors declare no conflicts of interest.

**Data availability.**

Data underlying the results presented in this paper are not publicly available at this time but may be obtained from the authors upon reasonable request.

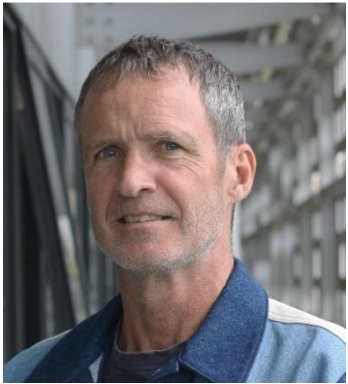

**Philippe Lalanne** is a CNRS research scientist at Institut d'Optique d'Aquitaine, specializing in nanoscale electrodynamics. He graduated at Ecole Normale Supérieure de Lyon in 1987. He has made several significant contributions, including the development of novel modal theories, the formulation of general principles for designing high-Q microcavities, and the clarification of the role of plasmons in extraordinary optical transmission phenomena. In the 1990s, he also pioneered the development of large-NA metalenses using high-index nanostructures. His current research interests focus on the non-Hermitian interactions between light and nanoresonators, as well as optical metasurfaces. He received several distinctions, including the prestigious 2023 ERC Advanced grant and the Grand Prix Léon Brillouin of the French Optical Society in 2024. He is a fellow of the Institute of Physics (IOP), SPIE and OPTICA.

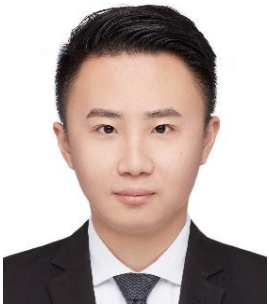

**Miao Chen** received the B.S. degree in microelectronics and solid state electronics from the University of Electronic Science and Technology of China, in 2016. He then reviced the Ph.D. degree in microelectronics and solid state electronics at Institute Of Semiconductors, Chinese Academy of Sciences, in 2022. He now works as a postdoctoral researcher at Laboratory for Photonics, Numerics and Nanosciences (LP2N), Institut d'Optique Graduate School. His research has so far been concerned with modelling of light scattering and light–matter interaction in disordered metasurface.

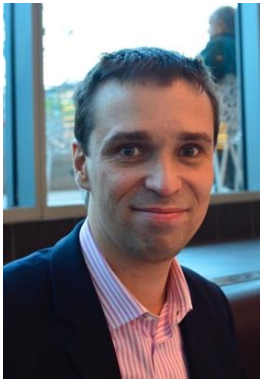

**Carsten Rockstuhl** received a Ph.D. from the University of Neuchâtel, Neuchâtel, Switzerland in 2004. After a Postdoc period at AIST in Tsukuba, Japan, he has been since 2005 with the Friedrich Schiller University of Jena, Jena, Germany. In 2013, he was appointed a full professor at the Karlsruhe Institute of Technology, Karlsruhe, Germany. He works on many aspects in the context of theoretical and computational nano-optics, where his latest efforts revolve around multi-scale modelling, inverse problems, scattering theory, and disordered photonic materials. He has served and serves the community as an editor with multiple journals. Moreover, he is a member of the Karlsruhe School of Optics and Photonics, where he currently acts as the dean of study, the Max Planck School of Photonics, and is a fellow of Optica.

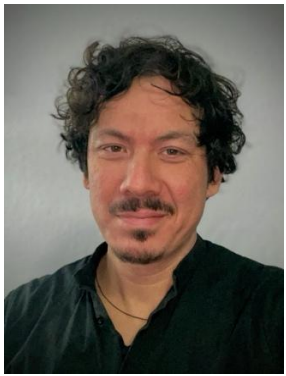

**Alexander Sprafke** is a researcher and group leader in the Institute of Physics at the Martin Luther University Halle-Wittenberg (MLU), Germany. He received a doctorate in physics from RWTH Aachen University, Germany, and held positions as a Visiting Research Fellow at the University of New South Wales, Sydney, Australia, and as a Guest Professor at the Lyon Institute of Nanotechnology, France. His research interest spans experimental nanophotonics and metasurfaces for solar energy applications as well as correlated-disorder materials including their scalable fabrication. He organized and co-chaired multiple OPTICA and SPIE conferences on Photonics for Solar Energy. In 2024, he was appointed Scientific Director of the Interdisciplinary Center for Material Science at MLU.

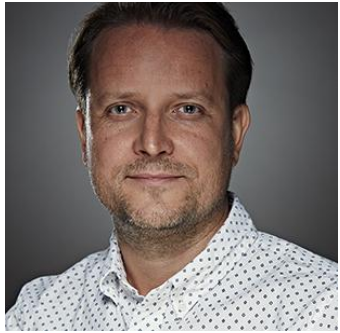
**Alexandre Dmitriev** earned his PhD in nanophysics in 2003 at Max-Planck-Institute for Solid State Research (Stuttgart, Germany) and EPFL (Lausanne, Switzerland). From mid-2004 he was at Chalmers University of Technology (Gothenburg, Sweden), first as EU Marie Curie Fellow, and later as Swedish Research Council assistant professor, and is currently full professor at the University of Gothenburg and the Head of Department of Physics. He was visiting professor at Stanford (USA) in 2016-2017 and 2018 and is a holder of the Swedish Foundation for Strategic Research Future Research Leader and Strategic Research Expedition awards (2010, 2019). He served as Chair (2010-2012) of one of formerly largest European research network in plasmonics (COST Plasmonics), was a Fellow of the Mobility for Regional Excellence of the Västra Götaland Region and Erskine Visiting Fellow at the University of Canterbury (New Zealand). His research interests are in experimental magnetophotonics, nanoplasmonics and nano-optics, biological and chemical sensing for sustainability, nano-photovoltaics, thermal management for energy efficiency.

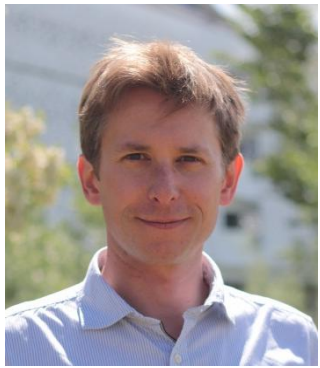
**Kevin Vynck** is a CNRS research scientist at the Institute of Light and Matter (iLM) in Lyon area (France), specialized in theoretical and numerical modelling light scattering and transport in complex nanostructures. He received his PhD from the University of Montpellier II in 2008 and his "habilitation" degree (HDR, in french) from the University of Bordeaux in 2023. Most notably, he contributed to the emergence of the concept of disorder engineering in optics and photonics, with various applications ranging from thin-film photovoltaics to visual appearance design. In 2019, he was awarded the CNRS Bronze Medal.